%% file: ms.tex
\documentstyle[epsfig,amsmath,amssymb,graphicx,soul,color,deluxetable,subfigure,lscape]{jaa}

\def\egs{\,erg~s$^{-1}$}      
\def\farcs{\hbox{$.\!\!^{\prime\prime}$}}

\def\gsim{\;\rlap{\lower 2.5pt\hbox{$\sim$}}\raise 1.5pt\hbox{$>$}\;}
\def\lsim{\;\rlap{\lower 2.5pt\hbo time lag between the formation x{$\sim$}}\raise 1.5pt\hbox{$<$}\;}
\def\la{\mathrel{\hbox{\rlap{\hbox{\lower4pt\hbox{$\sim$}}}\hbox{$<$}}}}
\def\ga{\mathrel{\hbox{\rlap{\hbox{\lower4pt\hbox{$\sim$}}}\hbox{$>$}}}}

\def\arcmin{\hbox{$^\prime$}}
\def\arcsec{\hbox{$^{\prime\prime}$}}

\def\farcs{\hbox{$.\!\!^{\prime\prime}$}}

\def\aj{AJ}
\def\apj{ApJ}
\def\aa{A\&A}
\def\mnras{MNRAS}

\begin{document}
\title[XMM-Newton view of eight young open star clusters]{X-ray observations of eight young open star clusters : I. Membership and X-ray Luminosity}
\author[Himali Bhatt et al.]{Himali Bhatt$^{1}$,
J. C. Pandey$^{2}$,
K. P. Singh$^{3}$,
Ram Sagar$^{2}$,
Brijesh Kumar$^{2}$ \\
$^{1}$Astrophysical Sciences Division, Bhabha Atomic Research Center, Trombay, Mumbai 400 085, India \\
$^{2}$Aryabhatta Research Institute of Observational Sciences, Manora Peak, Nainital 263 129, India\\
$^{3}$Tata Institute of Fundamental Research, Mumbai 400 005, India\\
}

\pubyear{xxxx}
\volume{xx}
\date{Received xxx; accepted xxx}
\maketitle
\label{firstpage}
\begin{abstract}
We present a detailed investigation of X-ray source contents of eight young open clusters with ages between 4 to 46 Myr using archival X-ray data from {\sc XMM-Newton}. The probable cluster memberships of the X-ray sources have been established on the basis of multi-wavelength archival data, and samples of 152 pre-main sequence (PMS) low mass ($<$ 2 M$_\odot$), 36 intermediate mass (2 - 10 M$_\odot$) and 16 massive ($>$ 10 M$_\odot$) stars have been generated. X-ray spectral analyses of high mass stars reveal the presence of high temperature plasma with temperature $<$2 keV, and mean $\rm{L_X/L_{bol}}$ of $\rm{10^{-6.9}}$. In the case of PMS low mass stars, the plasma temperatures have been found to be in  the range of 0.2 keV to 3 keV with a median value of $\sim$1.3 keV, with  no significant difference in plasma temperatures during their evolution from 4 to 46 Myr. The X-ray luminosity  distributions of the PMS low mass stars have been found to be similar in the young star clusters under study. This may suggest a  nearly uniform X-ray activity in the PMS low mass stars of ages $\sim$4 -- 14 Myr. These observed values of $\rm{L_X/L_{bol}}$ are found to have a mean value of  $\rm{10^{-3.6\pm0.4}}$, which is below the X-ray saturation level.The $\rm{L_X/L_{bol}}$ values for the PMS low mass stars are well correlated with their bolometric luminosities, that implies its dependence on the internal structure of the low mass stars. The difference between the X-ray luminosity distributions of the intermediate mass stars and the PMS low mass stars has not been found to be statistically significant. Their $\rm{L_X/L_{bol}}$ values, however have been found to be significantly different from each other with a confidence level greater than 99.999\% and the strength of X-ray activity in the intermediate mass stars is found to be lower compared to the low mass stars. However, the possibility of X-ray emission from the intermediate mass stars due to a low mass star in close proximity of the intermediate mass star can not be ruled out. 
\end{abstract}


\section{Introduction} \label{sec:st_int}

Young open star clusters constitute samples of stars of different masses with 
approximately the same age, distance and chemical composition, and 
these are homogeneous with respect to these properties. 
These clusters contain massive ($> 10$M$_\odot$), 
intermediate mass  (10 - 2 M$_\odot$) and PMS low mass ($<$ 2M$_\odot$) stars, and  
therefore, provide useful laboratories to study different mechanisms for the generation of 
X-rays in stars with different masses. In the massive stars,
the X-ray emission arises from shocks in radiatively-driven winds 
(Lucy \& White 1980; Owocki \& Cohen 1999; Kudritzki \& Puls 2000; Crowther 2007),
while  in the low-mass stars, rotation with convective
envelopes  drives a magnetic dynamo leading to strong X-ray emission (Vaiana et al. 1981; G\"{u}del 2004).
Intermediate mass stars, on the other hand, are expected to be X-ray dark because (a) the wind is not strong enough
to produce X-rays as in the case of  massive stars (see Lucy \& White 1980;
Kudritzki \& Puls 2000), and (b) being fully radiative internal structure, the dynamo 
action cannot support the X-ray emission. However, the mysterious
detection of X-rays from some intermediate mass stars still remains an
open question, and underlying physical mechanisms are not fully known (e.g., Stelzer et al. 2006). 

Further, the physical origin of X-ray emission from PMS low mass stars is also poorly understood. 
X-ray studies of low mass PMS stars in young clusters with ages
less than 5 Myrs like Orion, IC 348 and NGC 2264 (e.g., see Feigelson et al. 2003; Flaccomio et al. 2003;
Flaccomio, Micela \& Sciortino 2003; Stassun et al. 2004; Preibisch et al. 2005), and
in older zero-age-main sequence  (ZAMS) clusters like the Pleiades and IC 2391 with ages between 30 and 100 Myr 
(e.g., Micela et al. 1999; Jeffries et al. 2006; Scholz et al. 2007)
offer strong evidence that X-ray activity of PMS low mass stars originates due to coronal activity similar to 
that present in our Sun.
Studies of low mass stellar population with different ages, however, show an evolution of the X-ray activity levels in the
young stages ($<$5 Myr), the X-ray luminosity ($\rm{L_X}$) is in the range of $\rm{10^{29}-10^{31}}$ \egs, compared to
much lower activity seen in the older (ZAMS) stars, i.e., $\rm{L_X \sim{10^{29}}}$ \egs.
The X-ray activity is found to decay mildly with age during the evolution of PMS low mass stars from 0.1 to 10 Myr 
(Preibisch \& Feigelson 2005), while it steepens in the main sequence (MS) evolution from the ZAMS to a few Gyr 
age (Feigelson et al. 2004).   
Thus, the evolution of X-ray activity in the PMS stars is somewhat more
complicated than in the MS stars.
In addition, the ratio of X-ray luminosities to the bolometric 
luminosity ($\rm{L_X /L_{bol}}$) of PMS low mass stars in young clusters 
is found to be above the saturation level, i.e., $\rm{L_X /L_{bol}}\approx10^{-3}$, 
and uncorrelated with the rotation rates, while the low mass stars in ZAMS clusters 
show $\rm{L_X /L_{bol}}$  $\sim 10^{-8}$ -  $10^{-4}$.
The stellar X-ray activity  deviates from the saturation level for low mass stars 
in between 1 Myr to 100 Myr (Patten \& Simon 1996; G$\rm{\ddot{u}}$del 2004; Currie et al. 2009).
However, it is still not clear at which stage of the PMS evolution, 
the low mass stars deviate from the X-ray saturation level,
and which fundamental parameters govern their X-ray emission.

X-ray studies of clusters with intermediate age (5 to 30 Myr) have been few and far between.
An extensive study of young open clusters containing a number of stars with a range of masses
from massive to PMS low mass stars can address issues specific to the  mechanisms 
producing X-rays in stars with different masses. 
In addition, the young open clusters with a wide range of ages are also very 
useful targets for examining the evolution
of X-ray emission with age, especially in low mass stars.  
Multi-wavelength surveys of young open clusters provide an effective 
way to identify young cluster members among the huge number of foreground and background stars (a few Gyr) 
present in the same sky region, as young stars are more luminous in X-rays compared to the older field stars  
(e.g., Micela et al. 1985, Caillault \& Helfand 1985, Stern et al. 1981, Micela et al. 1988,1990, Preibisch et al. 2005).

The present work deals with characterizing the X-ray source contents of 
eight young open clusters with ages ranging from 4 to 46 Myr. 
This data sample bridge the gap between young clusters like the Orion
and the older clusters like the Pleiades, and constrain the evolution of X-ray emission with age for low mass stars. 
Samples of massive, intermediate and low mass PMS stars were collected using multi-wavelength archival data. 
The values of the extinction (E(B-V)), distances and age of the open clusters 
studied here are given in Table~\ref{tab:sample_info}.
The data were taken from {\sc XMM-Newton} pointed observations of the open clusters NGC 663, NGC 869, NGC 884 and IC 2602,
 whereas for the clusters NGC 7380, Berkeley 86 , Hogg 15 and Trumpler 18, data have been taken
from serendipitous observations  targeting  the massive stars HD 215835, V444 Cyg, WR 47 and supernova remnant SNR MSH11-62, respectively. X-ray emission characteristics of these eight young open clusters have been investigated 
here for the first time.
However,  X-ray emission from a few massive stars in the open clusters NGC 7380, Berkeley 86 , Hogg 15 and Trumpler 18 have been reported 
earlier (for details see \S\ref{sec:mass}).
In addition, previous spectral studies of the X-ray sources in the open cluster NGC 869 (h Persei) have been limited to
a region of size $\sim$ 15\arcmin (diameter) with {\sc Chandra} (Currie et al. 2009).
The present data cover the entire NGC 869 cluster region (28\arcmin)
due to the large field of view of the {\sc XMM-Newton}. 
The paper is organised as follows: 
the details of X-ray observations and data reduction procedure are presented in \S\ref{sec:observation}.
We have attempted to ascertain the cluster probable membership of 
 X-ray sources in \S\ref{sec:mem}.
X-ray variability and spectra of the cluster members are presented in \S\ref{sec:lc_spec} and \S\ref{sec:spec}, respectively.
The X-ray properties of cluster members are discussed in \S\ref{sec:spec_prop}  and
results are summarized in \S~\ref{sec:summary}.

\section{X-ray Observations and data reduction}
\label{sec:observation}
{\sc XMM-Newton} carries three co-aligned X-ray telescopes observing simultaneously, and
covering 30$^\prime$ $\times$ 30$^\prime$ region
of the sky. It consists three CCD-based detectors:
the PN CCD (Str$\ddot{u}$der et al. 2001) and the twin CCD detectors MOS1 and MOS2 (Turner et al. 2001).
EPIC has moderate spectral resolution ($\rm{\frac{E}{\delta E} \sim 20-50 }$) and 
an angular resolution\footnote{http://heasarc.gsfc.nasa.gov/docs/xmm/uhb/onaxisxraypsf.html} of 
4.5$^{\prime\prime}$, 6.0$^{\prime\prime}$ and 6.6$^{\prime\prime}$ for PN, MOS1 and MOS2 detectors, respectively. It together constitute the European Photon Imaging Camera (EPIC). 
We have analysed  archival X-ray data from {\sc XMM-Newton} observations of eight young open clusters and 
the journal of observations is given in Table~\ref{tab:xray_observations}.
All three EPIC detectors were active at the time of observations with full frame mode.
Data reduction followed  the standard procedures using the {\sc XMM-Newton}
Science Analysis System software (SAS version 10.0.0) with updated calibration
files.
Event files for MOS and PN detectors were generated by using
tasks {\sc emchain} and {\sc epchain} respectively, which allow
calibration, both in energy and astrometry, of the events registered in each CCD
chip and combine them in a single data file.
We limited our analysis to the energy band to 0.3 -- 7.5 keV because
data below 0.3 keV are mostly unrelated to bona-fide X-rays,
while above 7.5 keV only background counts are present,
for the kind of sources that we are interested in. Event list files were extracted
using the SAS task {\sc evselect}.
Data from the three cameras were individually screened for high background periods
and those time intervals were excluded where the total count rate (for
single events of energy above 10 keV) in the instruments
 exceed 0.35 and 1.0 $\rm{counts~s^{-1}}$ for the MOS and PN detectors, respectively.
The useful exposure times, i.e., sum of good time intervals, obtained after screening the high background periods for each cluster 
and corresponding to each detector used, are given in Table~\ref{tab:xray_observations}.

\subsection{Detection of X-ray point sources}
\label{sec:detection}
Detection of point sources is based on the SAS detection task {\sc edetect\_chain}, 
which is a chain script of various sub tasks 
(for details see XMM  documentation\footnote{http://xmm.esac.esa.int/sas/current/documentation/threads/src\_find\_thread.shtml}). 
First, the input images were built in two energy ranges, a soft band (0.3-2.0 keV)
 and a hard band (2.0-7.5 keV) for all three {\sc EPIC} detectors with a pixel size of 2\farcs0,
corresponding to a binsize of 40 pixels in the event file where each pixel size corresponds to 0\farcs05.
The task {\sc edetect\_chain} was then used simultaneously on these images. This task determined the source parameters (e.g., coordinates, count rates, hardness ratios, etc.) by means of 
simultaneous maximum likelihood psf (point spread function) fitting to the source count
distribution in the soft and the hard energy bands of each {\sc EPIC} instrument. 
A combined maximum likelihood value in all three instruments was taken to be greater than 10, corresponding to a 
false detection probability of $\rm{\approx 4.5 \times 10^{-5}}$. 
The output source lists from the individual {\sc EPIC} cameras in different energy bands were merged into a common list and  
the average values for the source positions with count rates were calculated.
The final output list was thus created giving source parameters for the soft and the hard energy bands 
along with the total energy band 0.3 -- 7.5 keV.
Spurious detections due to inter-chip gaps between CCDs, the hot pixels and 
the surroundings of bright point source regions have been removed by visual screening.
Finally, the number of X-ray sources detected in NGC 663, NGC 869,
NGC 884, NGC 7380, Berkeley 86, IC 2602, Hogg 15 and Trumpler 18
were 85, 183, 147, 88, 95, 95, 124 and 208, respectively. 
The estimated  positions of the all X-ray point sources along with their count rates  
in the total energy band of 0.3-7.5 keV 
are listed in Table~\ref{tab:detection}. 
Each source has been ascribed a unique identification number (ID) which is also given in  Table~\ref{tab:detection}.

The count rate of an X-ray source detected using {\sc edetect\_chain} task with  2$\sigma$ significance and lying within the cluster radius has been considered  as the
detection limit for each cluster. These detection limits in terms of count rates have been converted into fluxes limits using the  count conversion factors (CCFs)
used for low mass stars (see \S~\ref{sec:spec}) and corresponding X-ray luminosities have been tabulated in  Table~\ref{tab:Source_info} for each cluster. 

\subsection{Infrared counterparts of X-ray sources}
\label{sec:crossidentification}
X-ray point sources detected in the clusters were cross-identified with  NIR sources listed in  
 the Two-micron all sky survey (2MASS)  Point Source Catalog (PSC; Cutri et al. 2003). 
The X-ray counterparts in the 2MASS catalogue 
were then searched for within a radius of 10$^{\prime\prime}$,  
only those with a 'read flag' (representing uncertainties in their magnitude) 
value of 1 or 2 were retained.
In several cases, multiple counterparts are possible in the 2MASS PSC corresponding to an X-ray source and
the number of  multiple counterparts are given in column 9 (N) in  Table~\ref{tab:detection}.
In such cases, NIR sources that are closest to an X-ray source have been adopted as corresponding counterparts of that source.
The $JHK_S$ magnitudes, the positions of the X-ray sources from the center of the corresponding open cluster (see \S\ref{sec:mem}), 
and the offsets between the X-ray and the NIR positions of the NIR counterparts are given in  Table~\ref{tab:detection}.
It was thus found that only 70\%, 77\%, 70\%, 86\%, 94\%, 78\%, 85\% and 93\%
of the X-ray sources in the open clusters  NGC 663, NGC 869, NGC 884, NGC 7380, Berkeley 86, IC 2602, Hogg 15 and Trumpler 18, respectively, 
have 2MASS NIR counterparts.
Optical spectroscopic catalogues of stars from 
 Webda\footnote{http://www.univie.ac.at/webda/navigation.html}  and Vizier\footnote{ http://vizier.u-strasbg.fr/viz-bin/VizieR}
were used for the optical  identification of X-ray sources (see  Table~\ref{tab:detection}).

\section{Cluster Membership of X-ray Sources}
\label{sec:mem}
The X-ray sources with identifiable counterparts in the NIR band may not necessarily be members of their 
respective clusters. It is difficult to decide cluster membership of an individual X-ray source,
because the cluster population is contaminated by foreground and background stellar sources (Pizzolato et al. 2000), 
and extragalactic sources (Brandt \& Hasinger 2005).
In order to find which of the X-ray sources actually belong to a cluster, the approach given by Currie et al.(2010) has been adopted here. 
The step by step procedure used is given below.  

\subsection{Center and radius of the clusters}
\label{sec:radius}
The stellar population associated with young open clusters is still embedded in parent molecular clouds, due to which 
a large variation in extinction is found  within young open clusters.
The young stars embedded within high extinction regions of the cluster and hidden in the 
optical bands may be  visible in the NIR band.
Therefore,  NIR  data from the 2MASS PSC (Cutri et al. 2003) were used 
to  estimate  the center of these clusters and their extents rather than the optical data.
The center of a cluster was first taken to be an eye estimated center of the cluster and then refined as follows.
The average RA(J2000) and DEC(J2000) position of 2MASS stars having $K_S\leq14.3$ mag (99\% completeness limit in $K_S$ band) and lying within  1\arcmin~ radius was computed.
The average RA(J2000) and DEC(J2000) were reestimated by using this estimated value of the center of the cluster.
This iterative method was used until it converged to a constant value for the center of the open cluster (see Joshi et al. 2008 for details).
Typical error expected in locating the center by this method is $\sim$5\arcsec.
The estimated values of the center of each open cluster are given in Table~\ref{tab:Source_info}.
The positions of the centers estimated from the NIR data are consistent within 1\arcmin~ to that estimated from  the optical data by Dias et al. (2002).
Assuming spherical symmetry for the cluster, a projected radial stellar density profile of stars was constructed  and
the radius at which the stellar density is at the 3$\sigma$ level above the field star density was determined.
The field star densities were estimated from the 200 $\rm{arcmin^2}$ region which is nearly more than 0.5 degree 
away from the cluster regions.
The estimated values of  field star densities are 2.21$\pm$0.10, 1.95$\pm$0.10, 1.95$\pm$0.10, 2.86$\pm$0.12, 6.91$\pm$0.18 and 13.14$\pm$0.26 $\rm{stars~arcmin^{-2}}$
for the clusters NGC 663, NGC 869, NGC 884, NGC 7380, Berkeley 86 and Hogg 15, respectively.
 The estimated values of the radii of the open clusters are given in  Table~\ref{tab:Source_info}.
We have adopted the estimated radii, reported in Table~\ref{tab:Source_info}, as a measure of the extent of the open clusters.
Our estimates of the radii of the clusters are larger than that of the values given by Dias et al. (2002) using optical data except for the open cluster Berkeley 86.
However, these values are consistent with the values given by Pandey et al. (2005) and Currie et al. (2010) in the case of NGC 663 and NGC 869, respectively.
It is not possible to define the cluster extent in the case of IC 2602 and Trumpler 18 because 
the boundary where the stellar densities merge into field star densities is not clearly marked in the  radial density profiles.
This may be either due to the very large size of the cluster in the case of IC 2602, and very small size in
 the case of Trumpler 18, 
or the stars in the clusters may not be distributed in a spherical symmetry.
For further analysis, we  used the radii given in  Dias et al. (2002) catalogue for these two open clusters.
The projected distances of X-ray sources from the center of the respective clusters are given in  Table~\ref{tab:detection}.
The number of X-ray sources within the radius of cluster are also given in Table ~\ref{tab:Source_info}.   
All the X-ray sources with a counterpart in NIR  and falling within the adopted radius  of the corresponding  cluster have been considered for further analysis 
to check if they are members of that cluster.

\subsection{Color-magnitude diagram of X-ray sources with NIR counterparts}
Assigning cluster membership to X-ray sources is a difficult task.
It is, however, easier to check if an X-ray source lying 
within the cluster radius is not a member by using  color magnitude diagrams (CMDs).
In Figure~\ref{fig:cmd}, we plot the CMDs using the  2MASS $J$ magnitudes and $(J-H)$ colors 
of the sources selected in \S\ref{sec:radius} and lying within the cluster radii. 
We define the fiducial locus of cluster members for each open cluster by the  post-main sequence isochrones from Girardi et al. (2002)
and  PMS isochrones  from Siess et al. (2000)  according
to their ages, distances and mean reddening (see Table~\ref{tab:sample_info}).
These locii have been shown by dashed lines in Figure~\ref{fig:cmd}.
Width of each of the cluster locii has been determined by, (1) uncertainties in the determination of distance and age of the cluster, 
(2)  uncertainties in the photometric  2MASS $J$ and $H$ magnitudes of the sources which is higher at the fainter end,
 (3) dispersion in the reddening, and (4) binarity. Equal mass binaries may be up to 0.75 mag more 
luminous than single stars. 
The X-ray sources which are lying outside this fiducial locus  
for a given cluster are excluded from being members in that cluster.
The number of stars thus excluded from being members are 
10, 22, 11, 9, 2, 42, 14 and 2, in the open clusters
 NGC 663, NGC 869, NGC 884, NGC 7380, Berkeley 86, IC 2602, Hogg 15 and
Trumpler 18, respectively. 

The remaining X-ray sources  were further screened 
for probable membership of the respective open clusters.
Each  of these X-ray source was investigated on the basis of 
information given in the optical spectroscopic catalogues
from Vizier services. This spectroscopic information with references is given in  Table~\ref{tab:detection}. 
The sources for which the spectroscopic characteristics did not match with their 
photometric location in the CMDs were no longer considered for membership of the corresponding cluster, and thus
removed from the list of probable members.
 This method is useful for removing the foreground contamination. However, background contamination is very difficult to separate.
Therefore, we have further cross-identified these selected sources  in the  
all-sky comprehensive catalogue of radio and X-ray associations by Flesch (2010), in which the 
probability of a source being a quasi stellar object (QSO), a galaxy or a  star has been given.
Those sources  for which the probability for being a star is less than 20\% were also 
 removed from the list of selected sources.
Using this method, the number of additional  X-ray sources that are no longer considered as members of  
the open clusters  NGC 663, NGC 869, NGC 884, NGC 7380, Berkeley 86, IC 2602 and Hogg 15 were 5, 1, 3, 2 , 1, 2 and 1, respectively.

The X-ray sources that are no longer considered as members of a cluster are 
 marked by the symbol of a cross in 
Figure~\ref{fig:cmd} and listed as "N" in Table~\ref{tab:detection} (column 14).
The remaining X-ray sources are considered to be the probable members of their respective clusters.
We could thus assign probable cluster membership for 21, 70, 34, 25, 8, 10, 30 and 6 X-ray sources in 
clusters NGC 663, NGC 869, NGC 884, NGC 7380, Berkeley 86, IC 2602, Hogg 15 and Trumpler 18, respectively, and
listed as "Y" in Table~\ref{tab:detection} (column 14).
Further, proper-motion and/or spectroscopic studies are  needed to confirm the membership of specific X-ray sources.
However, proper motions at a distance of 2.0 kpc are extremely hard to detect.  
X-ray sources lying outside the cluster radius remain as unclassified.

\subsection{Mass estimation of X-ray stars}
The masses of X-ray stars, identified as probable members of the clusters, were estimated using 
 theoretical isochrones of Girardi et al. (2002) for MS stars and Siess et al. (2000) for PMS stars.
The boundaries corresponding to a 10$M_\odot$ (Massive) star, 10--2 $M_\odot$ (Intermediate mass) star
 and a 2 $M_\odot$ (Low mass) star were derived from the J magnitudes using model isochrones 
corrected for distance, age and reddening for each cluster, and 
are shown in  Figure~\ref{fig:cmd} by arrows. The estimated mass of each star identified as 
a probable member is given in  Table~\ref{tab:detection}. 
In Figure~\ref{fig:cmd}, the massive stars, intermediate mass stars and low mass  stars are
 marked by the symbols of star, triangle and  dots, respectively.

\section{Variability of X-ray sources }
\label{sec:lc_spec}
X-ray emission of stars is known to be variable, and before estimating their luminosity function it is
important to first study their variability.
Due to its highest sensitivity, data from the  PN detector of EPIC were used for variability and spectral analysis.
Light curves and spectra for all the probable members were extracted using circular extraction 
regions centered on the source position provided by {\sc edetect\_chain} task in the energy range of  0.3--7.5 keV.
X-ray sources either falling in the inter-chip gaps in the PN detector or having total counts below 40 in the PN detector were ignored
for variability and spectral analyses.
The wings of the psf for bright sources are often largely contaminated
by emission from neighboring sources, therefore, the radii of extraction regions
were varied between 8$^{\prime\prime}$ and 40$^{\prime\prime}$ depending on the position
of the source in the detector and
its angular separation with respect to the neighboring X-ray sources.
The background data were taken from several neighboring source-free regions on the detectors.
For the timing analysis, we have binned the data with 300 --  5000 s according to the count rate of the sources. 
Due to poor count statistics, there were several time intervals in which count rates were lesser than 5, therefore,
we were not able to perform the $\chi^2$--test for variability analysis.
Fractional root mean square (rms) variability amplitude ($\rm F_{var}$) was estimated to quantify the variability in the X-ray light curves. 
 The   $\rm F_{var}$ and  the error in $\rm F_{var}$ (\rm{$\sigma_{F_{var}}$} ) have been defined as follows (Edelson et al. 2002; Edelson, Krolik \& Pike 1990) and given in Table~\ref{tab:Int_lo_spec}.

\begin{equation}
\rm {F_{var}  = \frac{1}{<X>} \sqrt{S^2 - <\sigma^{2}_{err}>}} 
\label{eq:fvar}
\end{equation}

\begin{equation}
\rm  {\sigma_{F_{var}} = \frac{1}{F_{var}} \sqrt{\frac{1}{2N}}  \frac{S^2}{<X>^2}}
\label{eq:fvarerr}
\end{equation}

where $S^2$ is the total variance of the light curve, $<\sigma^{2}_{err}>$
is the mean error squared and $\rm{<X>}$ is the mean count rate, however, 
$\rm F_{var}$ can not be defined when $S^2$ is lesser than $<\sigma^{2}_{err}>$. 
$\rm F_{var}$ quantifies the amplitude of variability with respect to the mean count rate.
The $\rm F_{var}$ is found to be more than 3$\sigma$ of its error for seven sources (see Table~\ref{tab:Int_lo_spec}), therefore,
these sources are considered as variable.
The lightcurves of six sources show characteristics of flares and analyses of these flares are presented 
in Bhatt et al. (2013) (hereafter, Paper II).
The background subtracted lightcurve of one remaining source with ID \#20 in the cluster IC 2602  
with ID \#20 in the cluster IC 2602 is shown in Figure~\ref{fig:ic2602_20}.
This source is very close to V554 Car, which is classified as BY-Dra type variable 
(Kazarovets, Samus \& Durlevich 2001).  
The X-ray lightcurve of the source show $\sim$20\% of variability with respect to its mean count rate 
during observational time scale and does not show any flare-like feature.   
 BY Dra type of star may have rotational modulation in X-rays (see Patel et al. 2013).

\section{X-ray spectra}
\label{sec:spec}
Spectral characteristics of stars are also required before one can estimate their luminosity functions.
X-ray spectra of the sources with counts greater than 40 have been generated using the SAS task {\sc especget},
which also computed the photon redistribution matrix and  ancillary matrix.
For each source, the background spectrum was obtained from source-free regions
chosen according to the source location (same regions as used in the generation of light curves).
Spectral analysis was performed based on global fitting using
the Astrophysical Plasma Emission Code ({\sc{apec}}) version 1.10 modelled by Smith et al. (2001) and
 implemented in the {\sc xspec} version 12.3.0. The plasma model
{\sc{apec}} calculates both line and continuum emissivities for a hot,
optically thin plasma that is in collisional ionization equilibrium.
The absorption towards the stars by interstellar medium was accounted for by using
a multiplicative model {\sc phabs} in {\sc Xspec} which assumes
the photo-electric absorption cross sections according to
Baluci$\acute{n}$ska-Church \& McCammon (1992).

The simplest spectral model considered, is that of
an isothermal gas which we refer to as the "{\sc{1T apec}}" model. This
model is expressed as {\sc{phabs}} $\times$ {\sc{Apec}}.
We adopted the approach used in Currie et al. (2009) for X-ray spectral fitting  and
used an initial temperature kT of 1.5 keV to start the spectral fitting which is a
compromise between values typical of stars in younger clusters (e.g., M17; Broos et al. 2007) and stars in older clusters (e.g.,
the Pleiades; Daniel, Linsky \& Gagn$\acute{e}$ 2002). 
Elemental abundance parameter with a value of
0.3 solar is routinely found in fits of stellar X-ray spectra, and was thus fixed to this value
in our analysis (Feigelson et al. 2002; Currie et al. 2009) for intermediate and low mass stars. 
For massive stars, however, abundance
parameter of 0.2 solar was fixed for fitting (see Bhatt et al. 2010; Zhekov \& Palla 2007).
The value of absorption column density, $\rm{N_H}$, was fixed throughout the fitting 
to the value derived using the relation given by Vuong et al. (2003),  $\rm {N_H = 5 \times 10^{21}\times E(B-V) cm^{-2}}$, and given in  Table~\ref{tab:sample_info}.
The temperature, kT  and the normalization
were the free parameters in spectral fitting.
 We performed C-statistic model fitting  technique
rather than using  $\chi^{2}$-minimization technique because of the poor count statistics.
The temperature, normalization and unabsorbed flux values were derived by this fitting technique.
The estimated temperatures, EM and luminosities are given in Table~\ref{tab:massive} for the massive stars
and in  Table~\ref{tab:Int_lo_spec} for the intermediate and low mass stars. 
A few examples of X-ray spectra of massive, intermediate and 
low mass stars are shown in Figure~\ref{fig:spectra} along with 
the ratios of the X-ray data to the fitted model in the lower panels.

The X-ray fluxes of the probable cluster members having very poor count statistics (counts below 40) or
 which were lying between the inter-chip gaps between PN CCDs, 
were derived from their X-ray count rates in the EPIC detectors estimated from the SAS task {\sc {edetect\_chain}} 
(see \S\ref{sec:detection} and Table~\ref{tab:detection}). The CCFs to convert count rates into X-ray fluxes were
 estimated 
from WebPIMMS\footnote{http://heasarc.gsfc.nasa.gov/cgi-bin/Tools/w3pimms/pim\_adv} 
using {\sc 1T apec} plasma model. The value of model parameter $\rm{N_H}$ was
 fixed from Table~\ref{tab:sample_info} for respective clusters.
However, abundance parameter was fixed at  0.2 solar for massive stars (Zhekov \& Palla 2007; Bhatt et al. 2010), 
and  0.3 solar for intermediate and low mass stars (Feigelson et al. 2002; Currie et al. 2009).
The plasma temperature was fixed at 1.0 keV  for massive stars (N$\acute{a}$ze 2009). 
However, for intermediate and low mass stars, the plasma temperature  was taken as 
 the mean of the temperatures derived from the
spectral fitting of other bright stars in the cluster.
The mean values of X-ray temperatures of intermediate mass stars have been found to be 2.07, 1.30, 2.38 and 0.29 keV
for the open clusters NGC 869, NGC 884, NGC 7380 and  Hogg 15, respectively. In the case of low mass stars,
  the mean values of X-ray temperatures have been found to be 0.71, 2.05, 1.59, 2.80, 1.87, 1.06 and 0.97 keV for the open clusters
NGC 663, NGC 869, NGC 884, NGC 7380, Berkeley 86, IC 2602 and Hogg 15, respectively.
The derived values of conversion factors of count rates into unabsorbed fluxes for  massive, intermediate and low mass stars  have been given in the footnotes of
 Table~\ref{tab:massive} and Table~\ref{tab:Int_lo_spec}. 
For the sources either falling in between  the inter-chip gaps of PN CCDs or outside the PN coverage area,
the X-ray fluxes in the MOS1 and the MOS2 detectors were estimated from their count rates using CCFs in the MOS detector.
The average value of X-ray flux in the MOS1 and MOS2 detectors has been quoted in Table~\ref{tab:Int_lo_spec}. 
Thus, the X-ray luminosities were estimated from the derived values of the X-ray fluxes and given in
 Table~\ref{tab:massive} for massive stars and in Table~\ref{tab:Int_lo_spec} for intermediate and low mass stars.

X-ray spectrum of star \#79 in Berkeley 86 could not be fitted with the model used for 
spectral fitting, therefore, the  $\rm{N_H}$ parameter
was varied as a free parameter.  The best-fit value of  $\rm{N_H}$ has been found 
to be 3$\rm{\times10^{22}~cm^{-2}}$, which is 8 times higher than that expected in the direction of the open cluster
Berkeley 86 (see Table~\ref{tab:sample_info}). This points to either very high intrinsic 
extinction in the source or the source does not belong to the open cluster Berkeley 86.
For the stars showing flares, the values of parameters listed 
in Table~\ref{tab:Int_lo_spec}, were derived from the spectral fitting performed  
for their quiescent state data.

\section{ X-ray Properties of stars in different mass groups}
\label{sec:spec_prop}
X-ray spectral properties of massive, intermediate and low mass stars were analyzed separately, because
the production mechanism of X-rays are different for different types of stars.
The  bolometric luminosities ($\rm{L_{bol}}$) of the stars were derived from their bolometric magnitudes ($\rm{m_{bol}}$).
The absolute $J_0$ magnitudes were estimated from their observed 2MASS $J$ magnitudes using well constrained age, 
reddening and distance parameters of corresponding  open clusters in literature (see Table~\ref{tab:sample_info}).
The $\rm{m_{bol}}$ of the stars were derived from their  $J_0$ magnitudes by 
interpolating the $\rm{m_{bol}}$ between  $J_0$ magnitude points in theoretical isochrones
of Girardi et al. (2002) for MS stars and Siess (2000) for PMS stars,
depending upon the age of the open cluster.

\subsection{Massive stars}
\label{sec:mass}
Our sample contains 16 massive stars of which 6 were reported previously (see references in Table~\ref{tab:massive}).
The best fit spectral parameters for 8 stars are given in Table~\ref{tab:massive}. 
 The X-ray fluxes for four massive stars (see Table~\ref{tab:massive}) were derived from their count rates in PN detector by using CCFs. 
 The X-ray temperatures are found to be less than 1.2 keV in general.
However, X-ray temperatures are found to be higher in the case of high mass 
X-ray binary HD 110432 (Lopes de Oliveira et al. 2007)  
and [SHM202] 138.
The $\rm{L_X}$ of massive stars lie in the range of 10$^{31 - 35}$ \egs.
The $\rm{L_X /L_{bol}}$ for each massive star is derived and given in Table~\ref{tab:massive}.
The average value of $\rm{Log(L_X /L_{bol})}$ ~is found to be  -6.92 with standard deviation of 0.31. 
This value of $\rm{L_X /L_{bol}}$ ~is  broadly consistent with value derived for a sample 
of nearly 300 massive stars by N$\acute{a}$ze (2009). 

\subsection{Low mass stars}
Although, there is strong evidence that X-ray emission originates from magnetically confined coronal 
plasma in the PMS low mass stars (e.g., Preibisch et al. 2005), the relationship between rotation 
and X-ray activity in PMS low mass stars remained unclear. 
During the PMS phase, the low mass stars undergo substantial changes
in their internal structure, evolving from fully convective structure to a radiative core plus convective envelope structure.
Consequently, the stellar properties of low mass stars---$\rm{L_{bol}}$, magnetic activity and rotation etc.,
are also changing during the PMS phase.  
The dependence of $\rm{L_{bol}}$ and age upon 
X-ray emission is examined in the following sections.  

\subsubsection{X-ray temperatures} 
Most of these sources have plasma temperatures between 0.2 and 3 keV which
are consistent with values derived for PMS stars in young clusters e.g.,  NGC 1333 (Getman et al. 2002), 
Orion (Feigelson et al. 2002), NGC 1893 (Caramazza et al. 2012) and M16 (Guarcello et al. 2012).
The average plasma temperature of the stars in the open clusters appears
 to be constant for all stars undergoing PMS evolution from 4 Myr to 46 Myr, 
and the median value is found to be $\sim$1.3 keV.

\subsubsection{X-ray luminosity functions and their evolution with age }
\label{sec:xlf}
X-ray luminosity functions (XLFs) of low mass stars in different clusters have 
been derived using Kaplan Meier (KM) estimator of integral distribution functions 
and shown in  Figure~\ref{fig:lx_low}(a). No significant difference is observed 
in the XLFs of low mass stars with ages in the range of 4 to 14 Myr.
However the XLF of low mass stars 
in the open cluster IC 2602 with an age of 46 Myr appears  to be lower than that of others.
The mean values of $\rm{Log~L_X}$ with their standard deviations have been found to be 
31.26$\pm$0.38, 30.82$\pm$0.31, 30.81$\pm$0.26, 31.22$\pm$0.31, 31.01$\pm$0.18, 
29.10$\pm$0.65, 31.24$\pm$0.32 and 30.78$\pm$ 0.35 \egs~for the open clusters NGC 663, 
NGC 869, NGC 884, NGC 7380, Berkeley 86, IC 2602, Hogg 15 and Trumpler 18, respectively.
The mean values of $\rm{L_X}$ of low mass PMS stars are thus nearly similar 
in all the open clusters except IC 2602. 

The evolution of the mean value of $\rm{Log~L_X}$ with age is shown in Figure~\ref{fig:lx_low}(b).
The majority of the low mass stars in our sample have masses greater than $\rm{1.4~M_\odot}$ 
as seen in Figure~\ref{fig:cmd}, except for the stars in IC 2602.
The stars in the open cluster IC 2602 with masses above $\rm{1.4~M_\odot}$ may have
$\rm{L_X}$ below 27.57 \egs~(detection limits) and are not detected in the present study. 
It indicates a sudden decrease in the $\rm{L_X}$ between 14 to 46 Myr for the stars
with masses above $\rm{1.4~M_\odot}$. 
Thus $\rm{L_X}$ is nearly constant during the evolution of low mass stars in PMS phase from 4 to 14 Myr 
and may decrease thereafter.
Scholz et al. (2007) reported that the rotation rates increase in the first few Myr of their evolution.
It is, therefore, possible that  an increase in the X-ray surface flux
due to an increase in the rotation rate may be compensated
by a decrease in the stellar surface area during PMS evolution, between 1 to 10 Myr, as described by Preibisch (1997).
Between 10 to 40 Myr, the decrease in $\rm{L_X}$ may be linked with a  rapid 
spin down in the stars, as suggested by Bouvier et al. (1997). 
However, the faintest cluster members have not been detected here, therefore, 
the complete XLFs of these clusters cannot be derived since  
the mean luminosities of the entire cluster population may be lower than these values.

\subsubsection{X-ray to Bolometric Luminosity ratios}
\label{sec:lxlbol}
$\rm{L_X /L_{bol}}$ provides an estimate of the fraction of total stellar energy that is
dissipated through coronal heating, and the estimated values of the $\rm{L_X /L_{bol}}$ for low mass
stars in our sample are listed in Table~\ref{tab:Int_lo_spec}.
The mean values of $\rm{ Log (L_X / L_{bol}) }$ with standard deviations are found to be 
-3.86$\pm$0.41, -3.63$\pm$0.49, -3.63$\pm$0.51, 
-3.12$\pm$0.53, -4.00$\pm$0.84, -3.52$\pm$0.49, -4.02$\pm$0.78 and -3.85$\pm$1.28 for 
the open clusters  NGC 663, NGC 869, NGC 884, NGC 7380, Berkeley 86, IC 2602, Hogg 15 and Trumpler 18, respectively.
These values  have been found to be consistent with the values derived for the young clusters : the Orion (-3.39$\pm$0.63), 
IC 348 (-3.53$\pm$0.43) and NGC 2547 (-3.20$\pm$0.24) (see Alexander \& Preibisch 2012).
The derived values of mean $\rm{ Log (L_X / L_{bol}) }$ for each cluster are similar and 
the mean value $\rm{ Log (L_X / L_{bol}) }$ is found to be -3.6 with a standard deviation of 0.4 
for the collective sample of low mass stars within these open clusters.
 
The relation between $\rm{L_X}$ and  $\rm{L_{bol}}$ along with the isopleths of $\rm{ Log (L_X / L_{bol}) }$ = -3.0,-4.0 
are shown in Figure~\ref{fig:late_lx_lbol}(a). 
It can be seen that most of the sources have  $\rm{L_X / L_{bol} }$ values  below the saturation level. 
The distribution of $\rm{ Log (L_X / L_{bol}) }$ for all the low mass stars in all the clusters 
has been shown in Figure~\ref{fig:late_lx_lbol}(b) which is derived
using  the KM estimator of integral distribution functions. It shows that only 15\% of the X-ray sources have
$\rm{ L_X / L_{bol} }$ values above the saturation level.
There are five sources with $\rm{ Log (L_X / L_{bol}) }$ greater than -2.5; values 
that are very unlikely to be found for stellar sources. 
This is possibly the result of these sources not being members of the corresponding clusters,
therefore, their $\rm{L_X}$ and $\rm{L_{bol}}$ may have not been estimated properly. 
The $\rm{L_X}$ values of these few sources are marked with a symbol 
of question mark in Table~\ref{tab:Int_lo_spec}. These sources are marked with the symbol of cross in
Figure~\ref{fig:late_ratio_lbol}, and were not considered for further analysis.
The evolution of $\rm{ Log (L_X / L_{bol}) }$ with age is shown in Figure~\ref{fig:evol_lx_lbol} and found that
the $\rm{ Log (L_X / L_{bol}) }$ is nearly constant during 4 to 46 Myr.

Among the low mass PMS stars in the Orion, the 
median $\rm{L_X /L_{bol}}$ is 2-3 orders of magnitude greater, i.e.,  $\approx10^{-3}$,
than that found within ZAMS stars and therefore their fractional X-ray luminosities are "saturated".
Currie et al. (2009) show that the stars with masses $>1.5M_\odot$ deviate from X-ray saturation by $\approx$10-15 Myr. 
The present analysis indicates that most of the low mass PMS stars come out from the saturation limit earlier than 4-8 Myr, 
which is quite early as compared to the age described by Currie et al. (2009), i.e., 10-15 Myr.
The X-ray emission depends upon the magnetic dynamo that is the result of
a combination of turbulent convection and rotation within the convection zone. 
As a low mass star contracts onto the MS, its internal structure changes and its outer convective zones shrinks.
Therefore, the evolution of fractional X-ray luminosity with age might be due to either the change in the 
internal structure of a star or spin-down rotation of a star during the PMS phase, or both.
Alexander \& Preibisch (2012) showed that there was no correlation between
the $\rm{L_X / L_{bol}}$ and the rotation period.
They also found some rather slowly rotating  stars (period $>$ 10 days)
with very strong X-ray activity, and suggested that 
instead of rotation it is the change in the internal structure of PMS stars 
during the evolution which is likely to be responsible for the 
generation of magnetic dynamo and consequently the X-ray emission.

The dependence of $\rm{L_X/L_{bol}}$ on $\rm{L_{bol}}$ is shown in Figure~\ref{fig:late_ratio_lbol}.
The correlation  coefficients between  $\rm{L_X/L_{bol}}$ and $\rm{L_{bol}}$ have been derived using 
Pearson product-moment test and Kendall tau rank test, and their values are found to be
-0.65 and -0.58, respectively for all the low mass stars in the sample. 
Thus, the probability of no  correlation  between $\rm{L_X/L_{bol}}$ 
and $\rm{L_{bol}}$, i.e., the null hypothesis, 
is estimated to be 2.2$\times\rm{10^{-16}}$ from both the tests.

The linear regressions have been calculated using the least-squares Marquardt-Levenberg algorithm (Press et al. 1992) corresponding to the  following relation for all the low mass stars 
and shown by continuous line in  Figure~\ref{fig:late_ratio_lbol}, 

\begin{equation}
\rm Log~(L_X/L_{bol}) = -0.48(\pm 0.05) \times Log~(L_{bol})  + 12.95 (\pm 1.63)   
\label{eq:2to1.4a} 
\end{equation}

For the low mass stars with ages between 4 to 14 Myr and shown by dashed 
line in  Figure~\ref{fig:late_ratio_lbol}, 
\begin{equation}
\rm Log~(L_X/L_{bol}) = -0.83(\pm 0.05) \times Log~(L_{bol})  + 24.97 (\pm 1.70)   
\label{eq:2to1.4b} 
\end{equation}

For the low mass stars in the open cluster IC 2602 with ages 46 Myr and shown 
by dashed and dotted line in Figure~\ref{fig:late_ratio_lbol}, 
\begin{equation}
\rm Log~(L_X/L_{bol}) = -0.36(\pm 0.17) \times Log~(L_{bol})  + 8.26 (\pm 5.69)   
\label{eq:2to1.4c} 
\end{equation}

The equation (\ref{eq:2to1.4a}) shows a power-law dependence of the fractional 
X-ray luminosity on $\rm{L_{bol}}$ during 4 to 46 Myr.
The power-law indices are found to be different for  
stars with age of 4--14 Myr and the stars in the cluster IC 2602 
with age $\sim$46 Myr from equation(\ref{eq:2to1.4b}) and  equation(\ref{eq:2to1.4c}), respectively.
Prebisch et al. (2005) showed  $\rm{{L_X}\propto {L_{bol}}}$ for the stars in Orion which
implies $\rm{L_X / L_{bol} }$ is nearly constant at 1 Myr.
For  NEXXUS sample of nearby field stars (Schmitt \& Liefke 2004), 
Prebisch et al. (2005) found $\rm{{L_X}\propto {L_{bol}^{0.42}}}$, which implies that
$\rm{(L_X / L_{bol})\propto{L_{bol}^{-0.58}}}$. 
The values of power-law indices  of $\rm{{L_X}}$ and $\rm{L_X / L_{bol}}$  relation 
for  the open clusters
NGC 663, NGC 869, NGC 884, NGC 7380, Berkeley 86, IC 2602 and Hogg 15 
are derived to be
-0.7$\pm$0.2, -0.8$\pm$0.1, -1.1$\pm$0.1, -1.0$\pm$0.1,
-0.9$\pm$0.1, -0.4$\pm$0.2, -1.0$\pm$0.1, respectively.
It implies that the  $\rm{(L_X / L_{bol}) }$ depends upon  $\rm{L_{bol}}$ 
during 4 to 46 Myr and this dependence upon $\rm{L_{bol}}$ may be started earlier than 4 Myr.
As low mass stars evolve to MS, their effective temperatures eventually  increase 
and the depth of their convective envelopes reduce, therefore their  $\rm{L_{bol}}$ changes.
During 4 Myr to 46 Myr, the $\rm{L_{bol}}$ increases nearly three times (Siess et al. 2000)  
for low mass star with masses in the range of 1.4 - 2.0 $\rm{M_\odot}$. 
This increase in $\rm{L_{bol}}$ can produce a decrease of nearly one-third in       
$\rm{(L_X / L_{bol}) }$ which can give a decrease of nearly 0.5 dex in logarithmic scale.
Such a variation cannot be distinguished using present data because 
the standard deviation in $\rm{Log (L_X / L_{bol})}$ is comparable with the 
decrease of 0.5 dex.

\subsection{Intermediate mass stars}

A convincing and unique explanation for the generation of
X-ray emission from intermediate mass stars  has
not been forthcoming, despite abundant speculations about the
possible mechanisms.
The presence of magnetic field of the order of
a few hundred Gauss (Donati et al. 1997; 
Hubrig, Sch$\ddot{o}$ller \& Yudin 2004; Wade et al. 2005) has been detected in  these stars,
that can support a shear dynamo which may be responsible for
X-ray emission from intermediate mass stars as in the T-Tauri stars.
At the same time, the option of unresolved companions  is also
considered because the intermediate mass stars are more likely to be found in
binaries, i.e., companion hypothesis (Baines et al. 2006; Stelzer et al. 2006 and references therein).

The detection limits are in the range of  $10^{27.6}$ -  $10^{30.8}$\egs~ for different clusters as 
they are located at different distances from the earth. 
For making a sample of intermediate mass stars from different clusters, 
a highest detection limit of $\rm{Log~L_X}\approx10^{30.8}$ \egs~ among all clusters 
(see also Table~\ref{tab:Source_info}) was used,
which shows that a star with $\rm{Log~L_X>30.8}$ \egs~ could be detected in any of the clusters.
In this way, a total  of 27 intermediate-mass stars were identified and examined further.
The  XLFs of the low mass and intermediate mass stars having $\rm{Log~L_X>30.8}$ \egs~ were derived using the KM estimator of integral distribution functions.
A comparison of the X-ray luminosities  and fractional X-ray luminosities of 
low mass stars and intermediate mass stars in the
present sample is  shown in Figure~\ref{fig:inter_late_xlf}.
The results of two sample tests are given in  Table~\ref{tab:two_sample}.
The results of the Wilcoxon Rank Sum, Logrank, Peto and Peto Generalized Wilcoxon and Kolmogorov-Smirnov (KS)  statistical 
tests show that the X-ray luminosity distribution of intermediate mass stars is different from that of 
low mass stars with confidence of 
93\%, 98\%, 92\% and 77\%, respectively.
Therefore, the X-ray luminosities of both types of stars above this limit of $\rm{Log~L_X( >30.8)}$ \egs~ 
are not significantly different from each other.
Further, the $\rm{L_X/L_{bol}}$ ratio of intermediate mass stars and low mass stars are different,
 with  a confidence of greater than 99.999\% 
using these statistical tests.

Recently, Balona (2013) suggest the light variation due to rotation modulation caused by star-spots in nearly 875 A-type stars using Kepler's data.
If A-type stars have spots, then it is natural to expect a magnetic field, and therefore X-ray activity in intermediate mass stars.
The median values of  $\rm{Log(L_X/L_{bol})}$ are found to be -5.06 and
-3.41 for intermediate mass stars and low mass stars, respectively.
It implies that if the intermediate mass stars themselves produce X-rays, 
the strength of the X-ray activity is possibly weaker as compared to the low mass stars.  
However, the possibility of the X-ray emission from  a nearby low mass star cannot be ruled out
 here due to the poor spatial resolution data of  {\sc {XMM-Newton}}.

\section{Summary and Conclusions}
\label{sec:summary}
We have described  the X-ray source contents of eight young open clusters using the {\sc XMM-Newton} data.
These clusters have ages ranging from 4 Myr to 46 Myr and thus provide a link between the X-ray properties of young clusters like the Orion and older clusters like the Pleiades.
The association and  membership of these X-ray sources
with stars has been deduced using optical and NIR data. Overall 152 X-ray sources have been identified with low mass PMS stars, 36 with
 intermediate mass stars and 16 with massive stars. 
The main results are summarized below.
\begin{enumerate}
\item{The X-ray temperatures, luminosities  and  fractional X-ray luminosities of massive stars are consistent with the values reported previously 
in the literature for other massive stars.}
\item{The plasma temperatures are found to be in the range of 0.2 keV to 3 keV  with a median value of 1.3 keV for all low mass stars irrespective of their ages.} 
\item{The observed XLFs of low mass stars in the open clusters with ages from 4 to 14 Myr appear to be similar, which implies that
$\rm{L_X}$ is nearly constant during PMS evolution from 4 to 14 Myr. Therefore, the decrease in $\rm{L_X}$  of low mass stars may occur during 14 to 100 Myr.
Non-detection of X-rays from the stars above $\rm{1.4~M_\odot}$ in the open cluster IC 2602 may give an indication of a sudden decrease in their $\rm{L_X}$  during 14 to 46 Myr.}
\item{$\rm{ Log (L_X / L_{bol}) }$ of most of the low mass stars are below the saturation limits and the mean value has been found to 
be -3.6 with a standard deviation of 0.4. This value is consistent with the values derived for other young clusters the Orion, IC 348 and NGC 2547. Thus, a deviation of low mass stars with masses greater than 1.4 $\rm{M_\odot}$ from X-ray saturation may occur before the age of 4-8 Myr, earlier than the age  derived by Currie et al. (2009), i.e., 10-15 Myr.}
\item{$\rm{(L_X / L_{bol}) }$ of low mass stars correlate well with their $\rm{L_{bol}}$, suggesting
its dependence on the internal structure of stars.}
\item{No statistically significant difference in $\rm{L_X}$ from the intermediate mass and the low mass PMS stars has been detected. 
But the observed $\rm{L_X /L_{bol}}$ for intermediate mass stars have been found to be significantly lower than that of low mass stars.
It possibly indicates that the strength of X-ray activity in intermediate mass stars is weaker than in the low mass stars.
Another possibility is that the origin of X-ray emission from intermediate mass stars might be the result of X-ray emission coming from an unresolved 
nearby low mass PMS star. Deeper and higher spatial resolution data with {\sc Chandra} is needed 
to check for this possibility and to estimate the complete XLFs of these clusters.}
\end{enumerate}


\section*{Acknowledgments}
We thank to an anonymous referee for his/her constructive comments.
This publication makes use of data from the Two Micron All Sky Survey, which is a joint project of
the University of Massachusetts and the Infrared Processing and Analysis Center/California
Institute of Technology, funded by the National Aeronautics and Space Administration and
the National Science Foundation, and data products from XMM-Newton archives using the high 
energy astrophysics science archive research center 
which is established at Goddard by NASA. 
We acknowledge XMM-Newton Help Desk for their remarkable support in X-ray data analysis.  
Data from Simbad, VizieR catalogue access tool, CDS, Strasbourg, France have also been used. 
HB is thankful for the financial support for this work through the INSPIRE faculty fellowship granted by the Department of Science \& Technology  India, and
 R. C. Rannot,  Nilesh Chouhan and  R. Koul for their support to finish this work.

\clearpage

\input{ref.tex}
\clearpage


\clearpage
\input{fig.tex}

\clearpage

  \input{sample_info.tex}

  \input{xray_observations.tex}

  \input{xray_iden_spec_sam.tex}

\clearpage

  \input{Cluater_Sources.tex}

  \input{Massive_new1.tex}

  \input{spectral_properties_new1.tex}

  \input{two_sample_test_new.tex}

\end{document}

%% file: fig.tex
\clearpage
\begin{figure*}
\vspace{1cm}{ 
\vspace{1.5cm}{
\includegraphics[width=1.6in,height=1.3in]{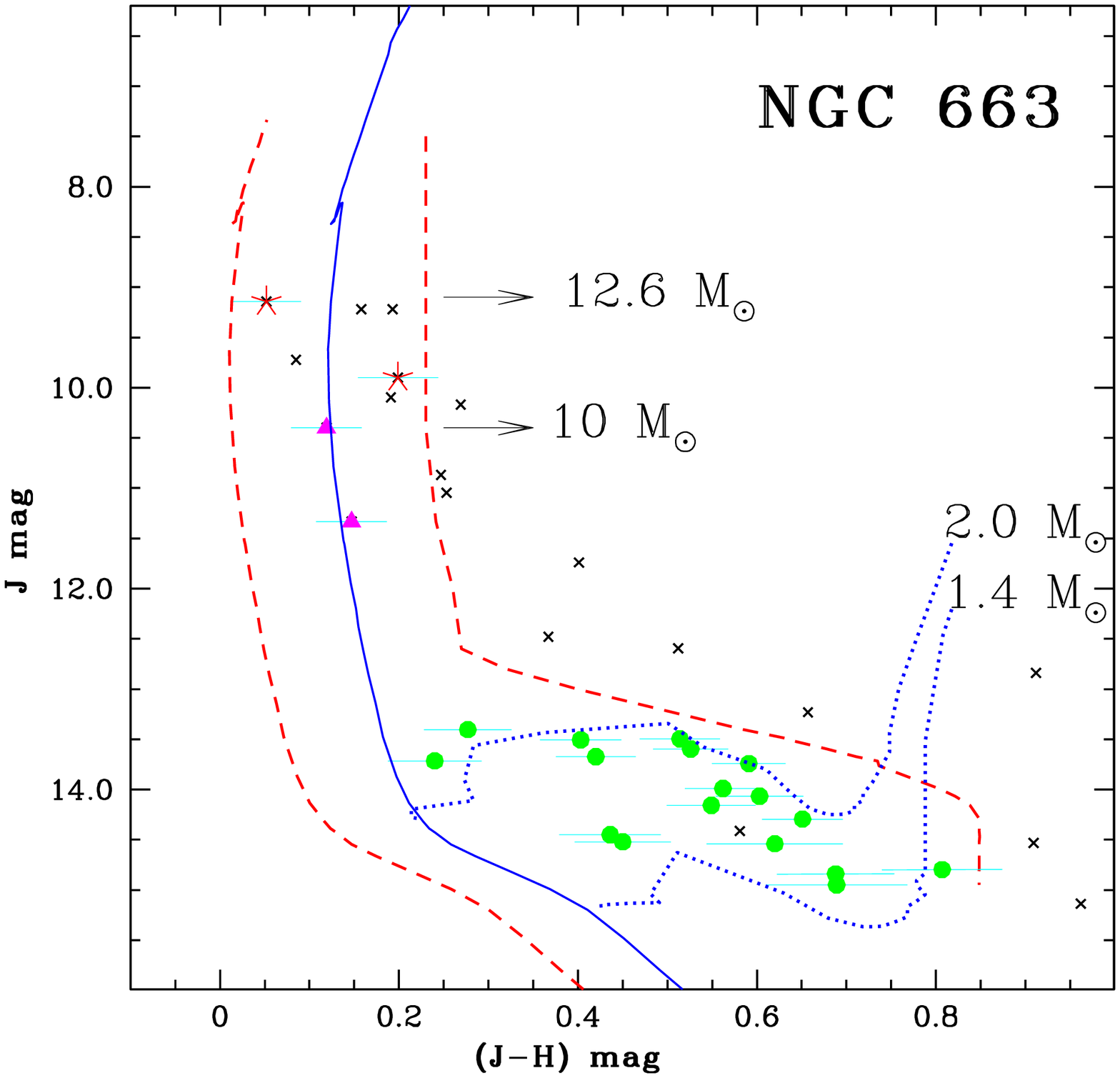}
\includegraphics[width=1.6in,height=1.3in]{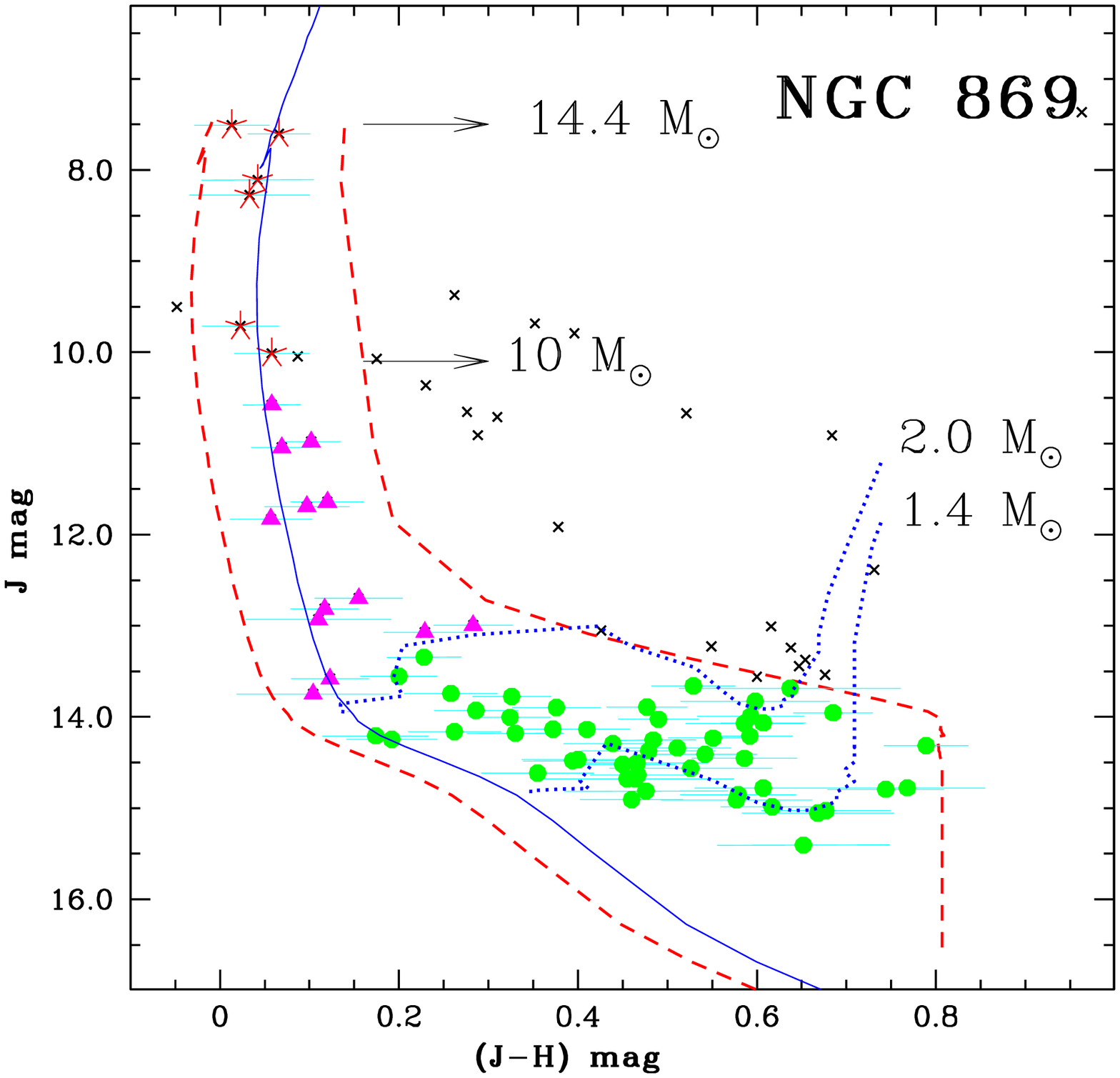}
\includegraphics[width=1.6in,height=1.3in]{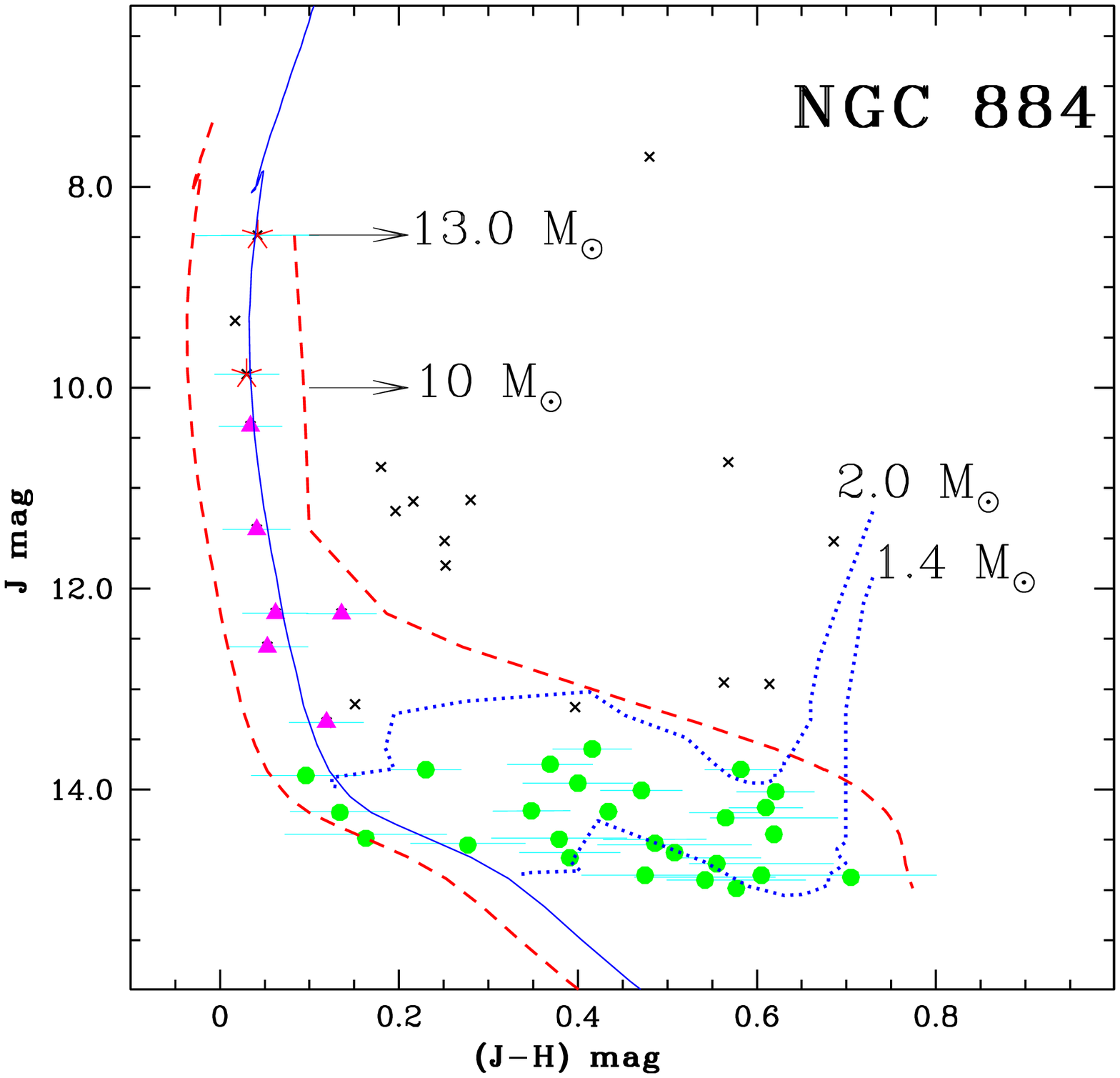}
}
\vspace{1.5cm}{
\includegraphics[width=1.6in,height=1.3in]{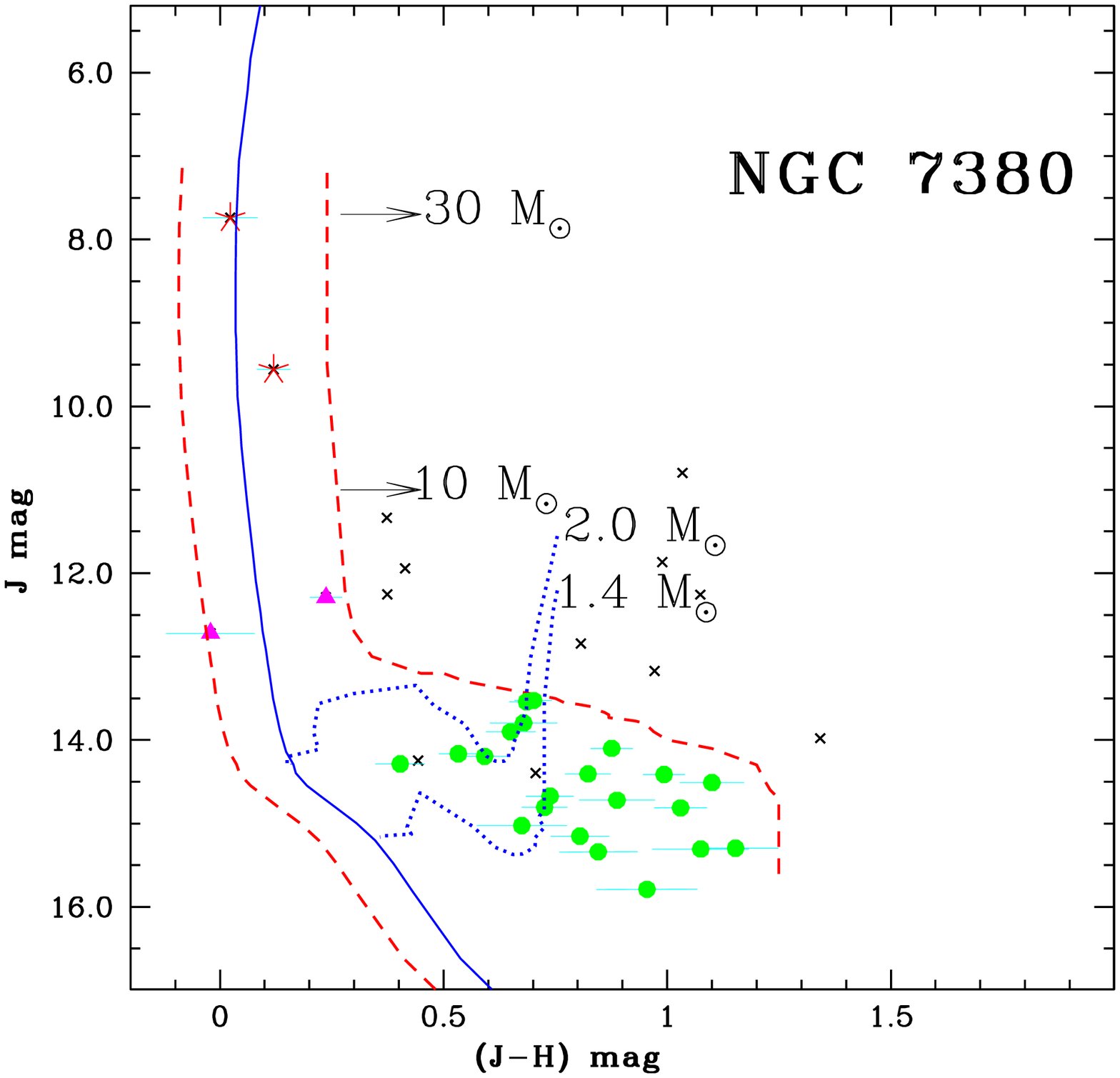}
\includegraphics[width=1.6in,height=1.3in]{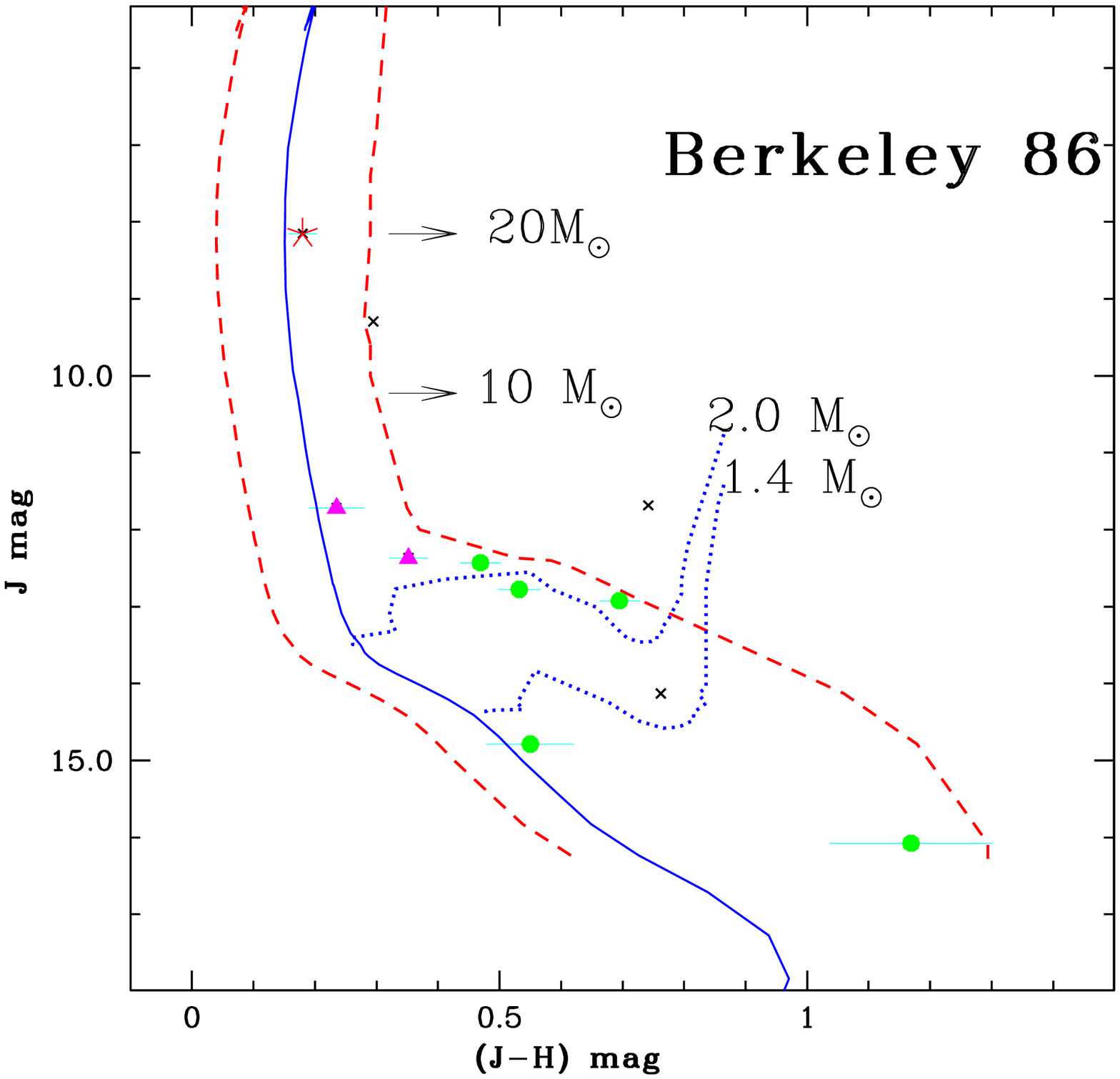}
\includegraphics[width=1.6in,height=1.3in]{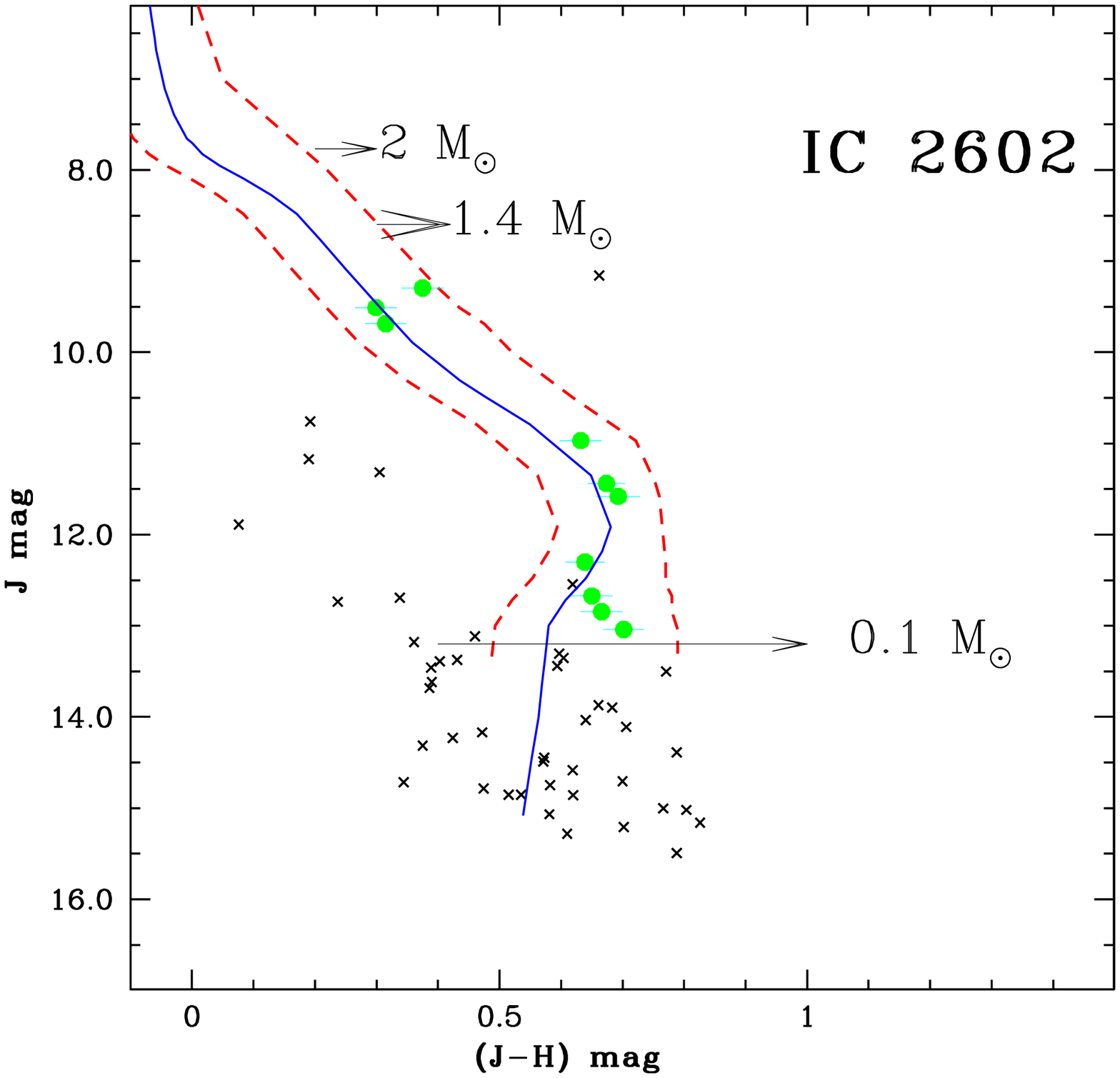}
}
\vspace{1.5cm}{
\includegraphics[width=1.6in,height=1.3in]{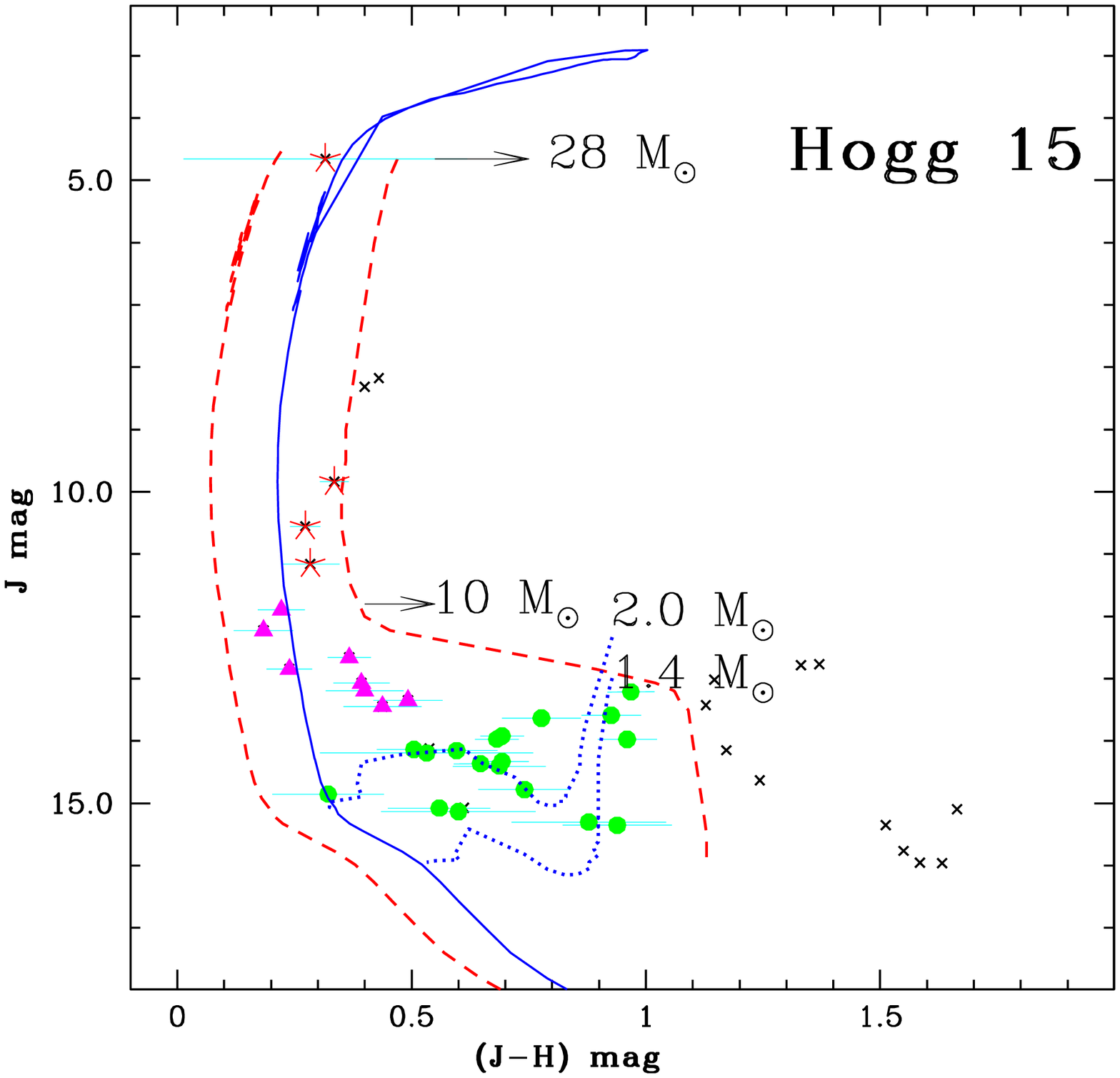}
\includegraphics[width=1.6in,height=1.3in]{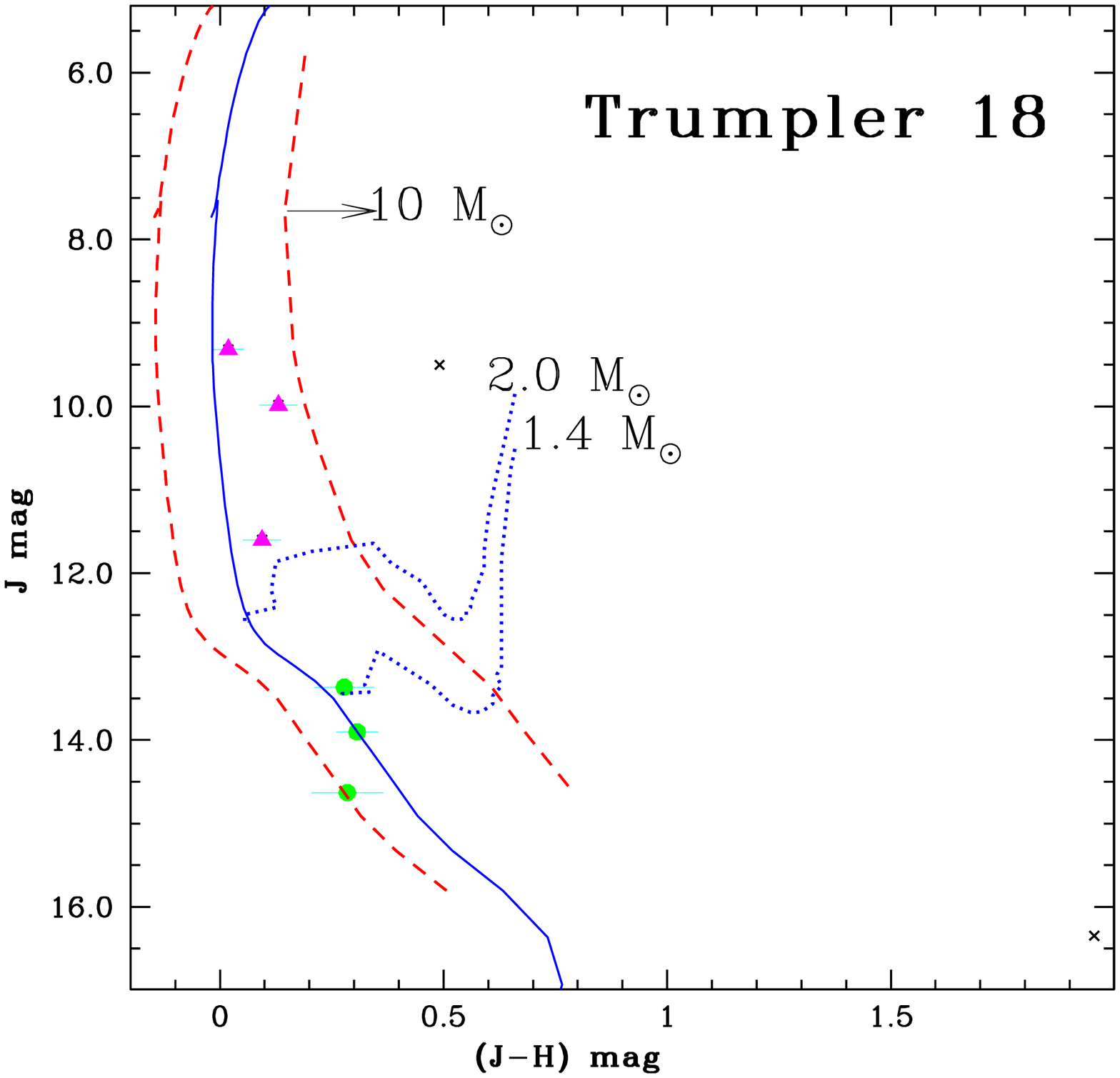}
}
}
\caption{$J$ versus $(J-H)$ color magnitude diagram (CMD) of the X-ray sources within cluster radius.
Post-main sequence  isochrones from Girardi et al. (2002) and PMS isochrones from Siess et al. (2000) are shown by solid and dotted lines (in blue).
The width of the cluster locus due to uncertainties in determination of distance, 
age, magnitudes and binarity
is shown by dashed lines (in red). The symbols of star, triangle and dots represent 
massive, intermediate and low mass stars, respectively, and
these boundaries are marked by arrows. The X-ray sources identified as a 
non-member of their respective open clusters are marked by the symbol of a cross. 
 }
\label{fig:cmd}
\end{figure*}

\begin{figure*}
\includegraphics[width=5.0in]{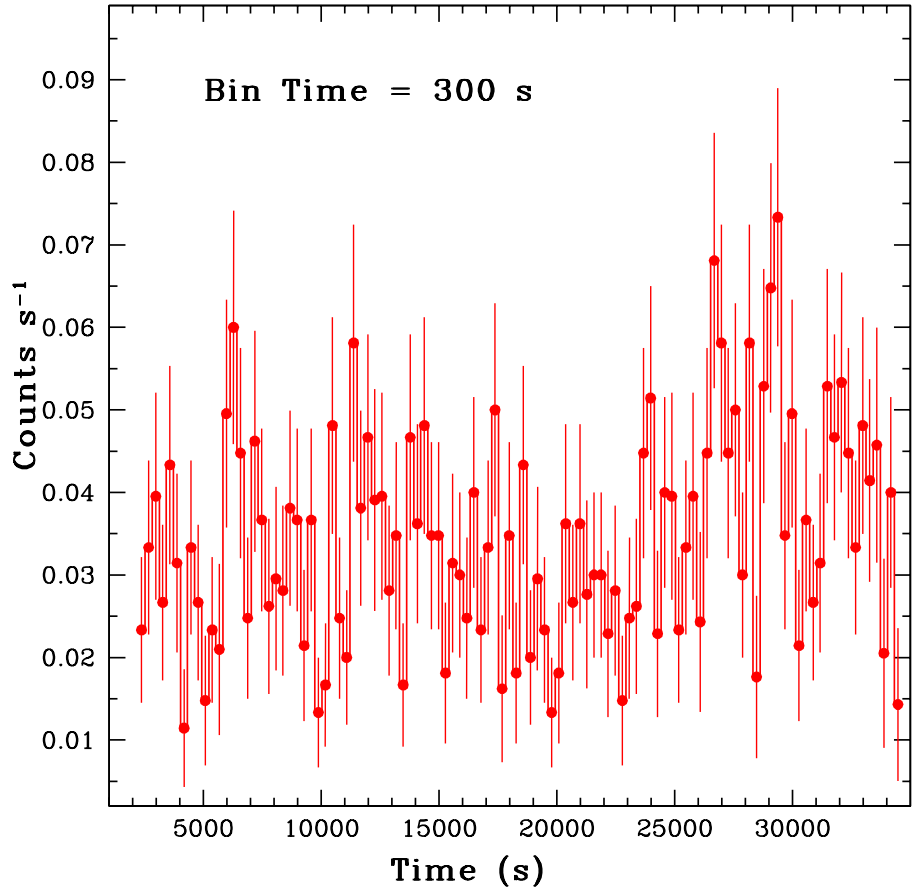}
\caption{Background subtracted X-ray lightcurve of the source with ID \#20 in the open cluster IC 2602. This source is close to the
BY Dra type variable star V554 Car.}
  \label{fig:ic2602_20}
\end{figure*}

\begin{figure*}
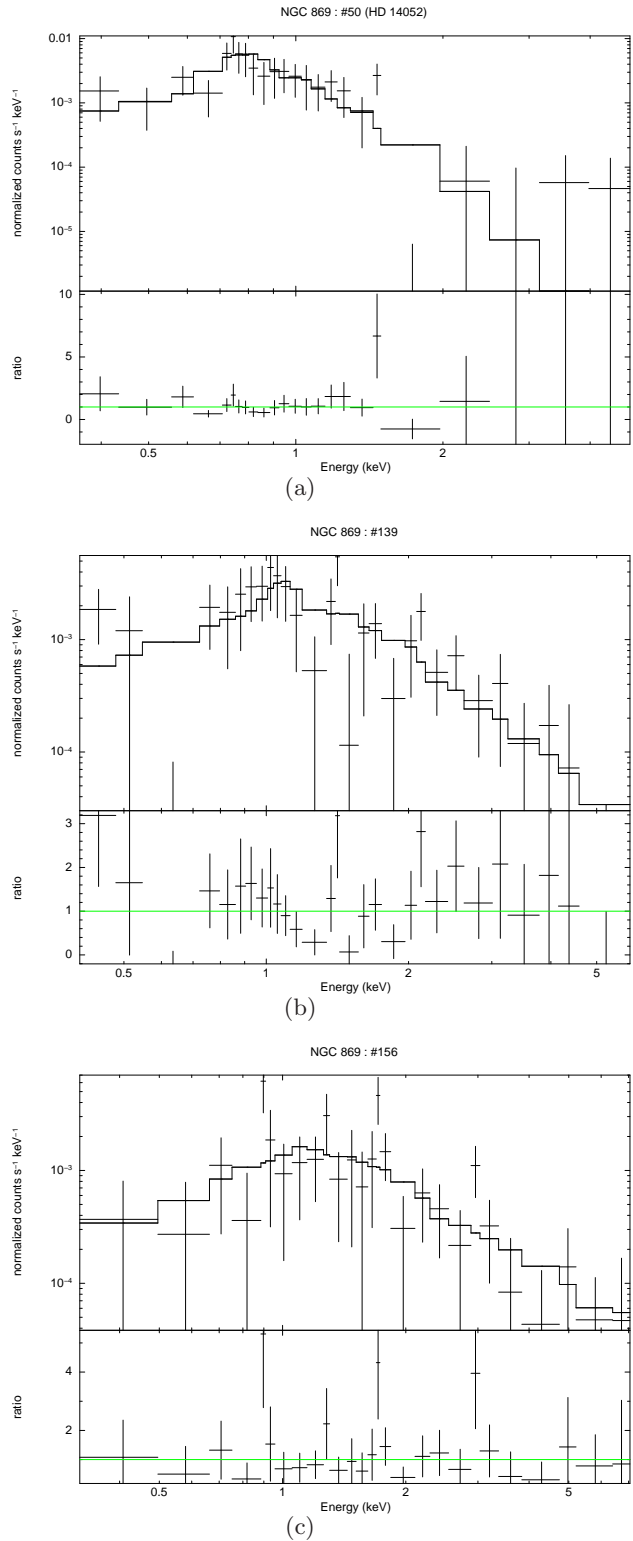

\subfigure[]{\includegraphics[width=2.5in,angle=270]{NGC869_50_HD14052.ps}}
\subfigure[]{\includegraphics[width=2.5in,angle=270]{NGC869_139_new.ps}}
\subfigure[]{\includegraphics[width=2.5in,angle=270]{source_156_spec.ps}}
\caption{A few example of X-ray Spectra of (a) Massive star, (b) Intermediate mass star and (c) Low mass star. The ID of the star with the information of its respective cluster is 
given at the top of each panel.}
  \label{fig:spectra}
\end{figure*}


\begin{figure*}
\subfigure[]{\includegraphics[width=4.0in, height=4.0in]{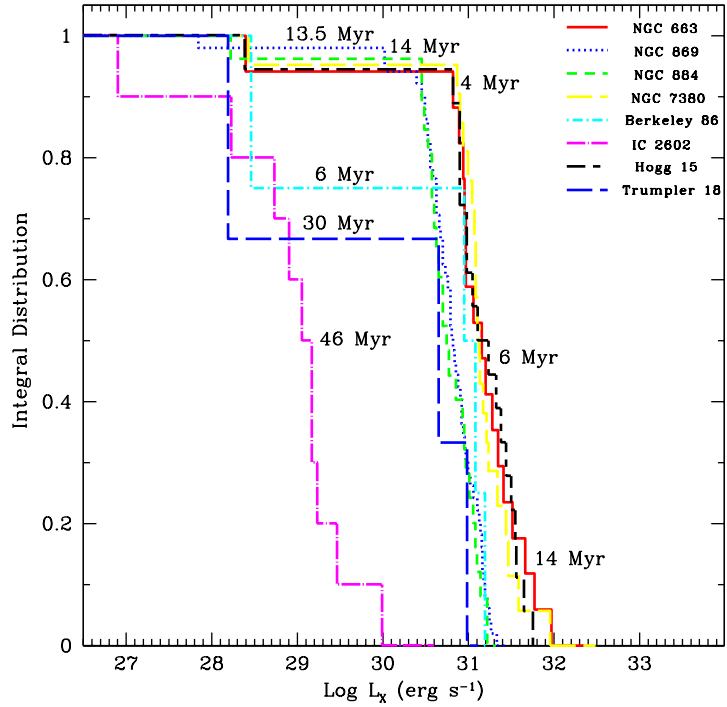}}
\subfigure[]{\includegraphics[width=4.0in, height=4.0in]{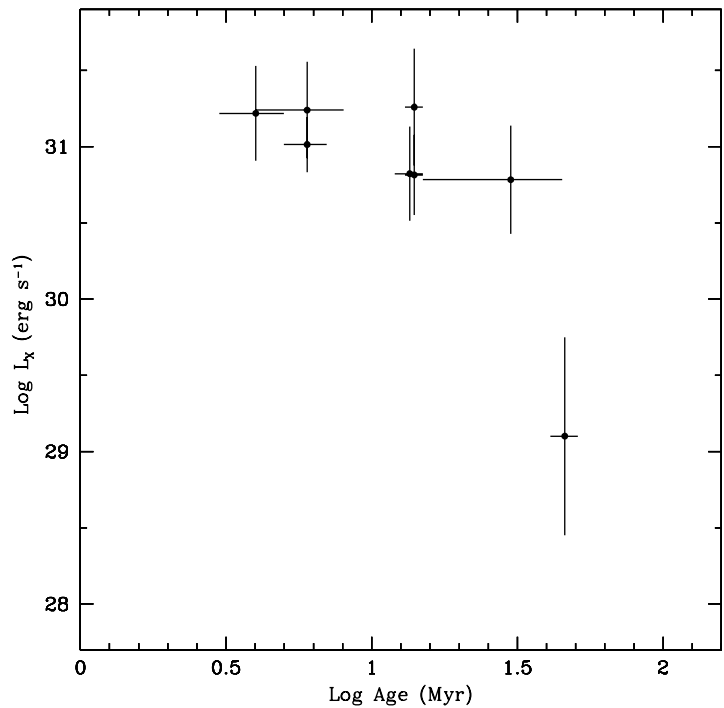}}
\caption{X-ray luminosities of low mass stars.(a) XLFs of low mass stars in different clusters. (b) Evolution of mean $\rm{L_X }$ of the clusters with age.    
}
\label{fig:lx_low} 
\end{figure*}

\begin{figure*}
\subfigure[]{\includegraphics[width=3.5in, height=3.5in]{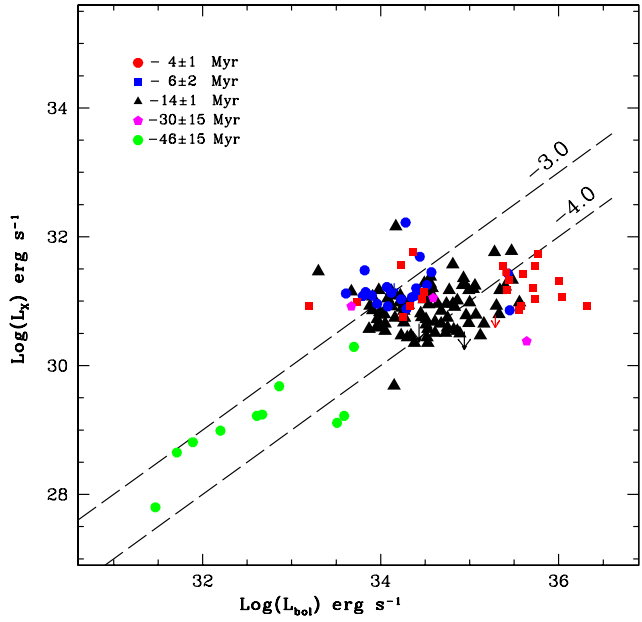}}
\subfigure[]{\includegraphics[width=3.5in, height=3.5in]{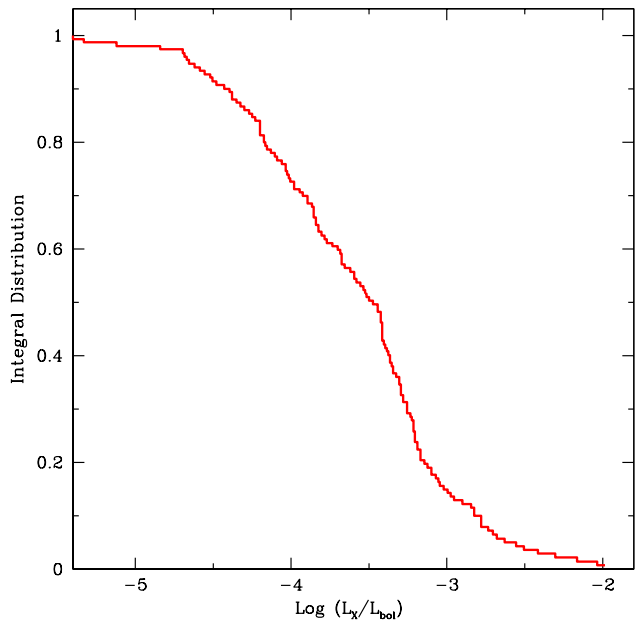}}
\caption{(a) Relation between $\rm{L_X}$ and $\rm{L_{bol}}$ for low mass stars in the sample.  
Dashed lines in each plot represent the isopleths of $\rm{ Log (L_X / L_{bol}) }$ and the values are given at above the each line.
(b) Distribution of $\rm{Log~L_X/L_{bol}}$ ratio for all the low mass stars in all the clusters.
}
  \label{fig:late_lx_lbol}
\end{figure*}

\begin{figure*}
\includegraphics[width=4.0in, height=4.0in]{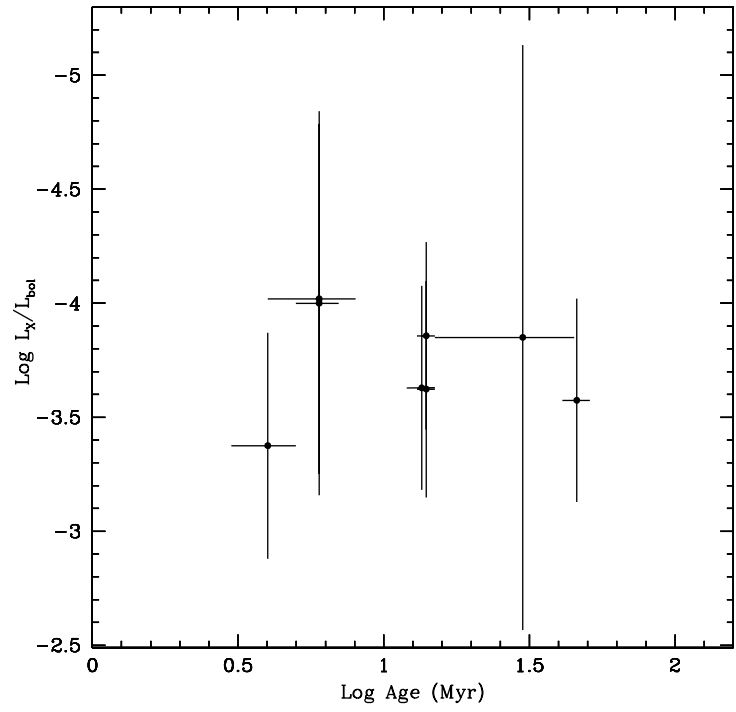}
\caption{ Evolution of mean $\rm{L_X/L_{bol}}$ of the clusters with age.    
}
  \label{fig:evol_lx_lbol}
\end{figure*}

\begin{figure*}
\includegraphics[width=4.0in, height=4.0in]{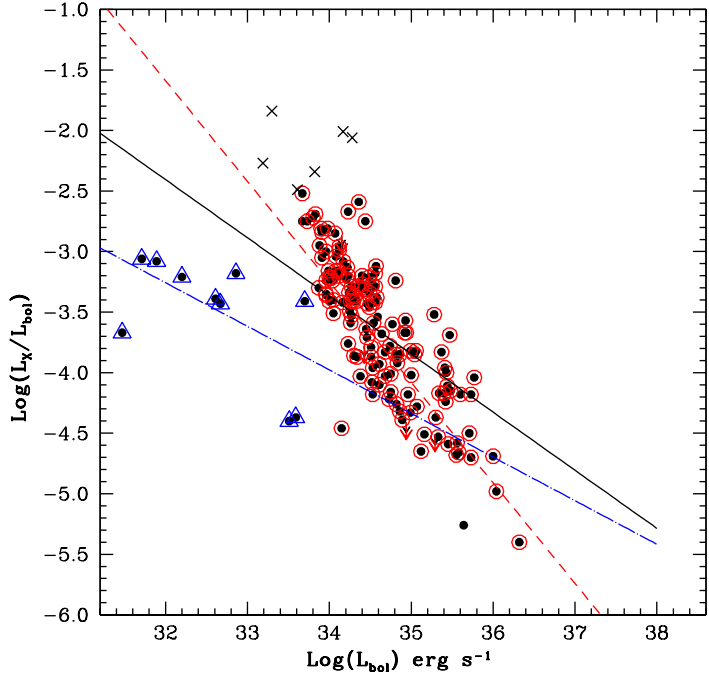}
\caption{Relation between $\rm{(L_X / L_{bol}) }$ and $\rm{L_{bol}}$  for low mass stars in the sample (dots),
for stars with age from 4 to 14 Myr (open circles), and for the stars in the cluster IC 2602 with age of 46 Myr (open 
triangles) derived using least square fitting and shown by continuous line, dashed line and dashed plus dotted line, respectively.
The stars with $\rm{ Log (L_X / L_{bol}) }$ above -2.5 have not been 
considered while deriving these relations and are marked by the symbol of cross.
}
 \label{fig:late_ratio_lbol}
\end{figure*}

 \begin{figure*}
  \includegraphics[height=5.0in]{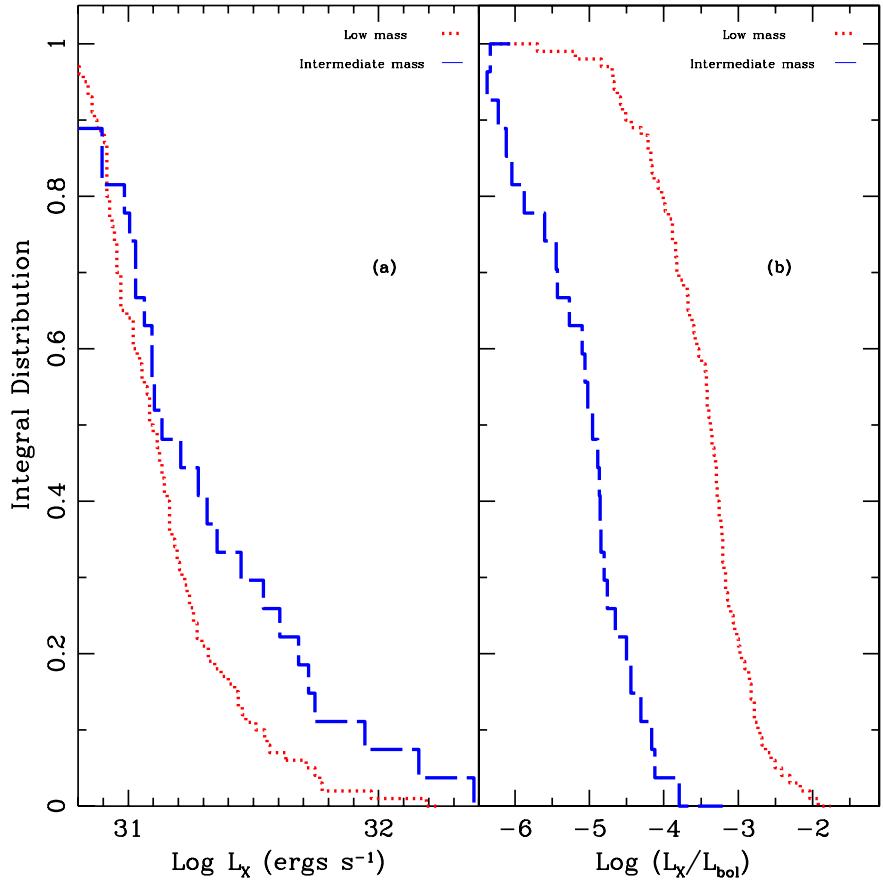}
  \caption{Comparison of the X-ray activity of  Low mass and Intermediate mass stars having $\rm{Log~L_X > 30.8~erg~s^{-1}}$ based on the Kaplan Meier Estimator. 
(a) distribution of $\rm{L_X}$ (b) distribution of $\rm{L_X/L_{bol}}$ ratio.}
  \label{fig:inter_late_xlf}
  \end{figure*}

%% file: sample_info.tex
\begin{table}
\scriptsize
\caption{The sample of the clusters under investigation with their basic parameters.}
\label{tab:sample_info}      
\begin{tabular}{c c c c c l}
\hline
Cluster Name  & E(B-V) &   $\rm{N_H}{^\P}$             &  Distance         & Age           & References   \\
              &  mag     & $10^{20}$ $\rm{cm^{-2}}$    &    pc             &  Myr          &             \\  
\hline
 NGC 663      & 0.80$\pm$0.15  &  40$\pm$7.5   &  2400$\pm$120   &  14$\pm$1     & Pandey et al. (2005)   \\ 
 NGC 869      & 0.55$\pm$0.10  &  28$\pm$5     &  2300$\pm$100   &  13.5$\pm$1.5 & Currie et al. (2010)   \\
 NGC 884      & 0.52$\pm$0.10  &  26$\pm$5     &  2300$\pm$100   &  14$\pm$1     & Currie et al. (2010) \\
 NGC 7380     & 0.60$\pm$0.10  &  30$\pm$5     &  2600$\pm$400   &   4$\pm$1     & Chen et al. (2011)\\
 Berkeley 86  & 0.95$\pm$0.10  &  47.5$\pm$5   &  1585$\pm$160   &    6$\pm$1    & Bhavya, Mathew \& Subramaniam (2007)\\
 IC 2602      & 0.035$\pm$0.01 & 1.75$\pm$0.5  &  150$\pm$2      &  46$\pm$5     & Dobbie, Lodieu \& Sharp (2010)\\
 Hogg 15      & 1.15$\pm$0.1   &  57.5$\pm$5   &  3000$\pm$300   &   6$\pm$2     & Sagar et al. (2001)\\
 Trumpler 18  &  0.3$\pm$0.04  &  15$\pm$0.2   &  1300$\pm$100   &  30$\pm$15    & Delgado, Alfaro \& Yun (2007)  \\
\hline
\end{tabular}
\newline
\noindent
$^\P$ : $\rm{N_H}$ is derived using the relation $\rm {N_H = 5 \times 10^{21}\times E(B-V) cm^{-2}}$ from Vuong et al. (2003).\\ 
\end{table}

%% file: xray_observations.tex
\begin{table*}
\scriptsize
\caption{Journal Of XMM-Newton Observations of eight young clusters.}             
\label{tab:xray_observations}      
\begin{tabular}{cccccccccc}       
\hline 
Cluster Name  & Observation  & Exposure  &~~~~~~~~~ Start time                &\multicolumn{3}{c} {Final retained exposure time}    & EPIC  & Offset from  \\

              &     ID       &  Time     & ~~~~~~~~UT               &    MOS1       & MOS2    & PN     & filter        &   Target \\           

              &               &  (sec)   &  (hh:mm:ss)             &\multicolumn{3}{c}{(ks)}          &         &   (arcmin)      \\
\hline
 NGC 663     &   0201160101   &   41915  &14 Jan 2004 22:40:26    & 32.59&  33.13&       28.69        & Medium  &1.606 \\
 NGC 869     &   0201160201   &   39509  &19 Jan 2004 04:39:50    & 38.50&  38.32&       34.88        & Medium  &1.871 \\                          
 NGC 884     &   0201160301   &   40620  &04 Feb 2004 15:13:25    & 33.08&  33.73&       26.08        & Medium  &1.013   \\                      
 NGC 7380    &   0205650101   &   31413  &19 Dec 2003 02:02:12    & 25.23&  25.46&       19.52        &  Thick  &3.621   \\                      
 Berkeley 86 &   0206240801   &   19921  &27 Oct 2004 23:25:39    & 18.86&  18.79&       15.11        &  Thick  &8.807    \\
 IC 2602     &   0101440201   &   44325  &13 Aug 2002 05:10:42    & 36.47&  36.69&       31.87        & Medium  &5.550      \\                    
 Hogg 15     &   0109480101   &  53040  &03 July 2002 15:51:56    & 48.30& 52.08 &       51.80        & Thick&4.509\\ 
 Trumpler 18$^{\dagger\dagger}$ &   0051550101   &  40822  &06 Feb  2002 01:13:19 &      &  35.85  & 39.20 & Medium&3.968\\ 
\hline
\end{tabular}
\newline
\noindent
$\dagger\dagger$ : The observations have not done in Prime Full window mode for MOS1 detector.
\end{table*}

%% file: xray_iden_spec_sam.tex
\begin{landscape}
\begin{deluxetable}{cccccccccccccccc}
\tabletypesize{\tiny}
\tablecaption{\small  X-ray information of all X-ray sources detected in detection procedure (for details see \S2.1) within the field of view of {\sc Xmm-newton} 
 along with their cross-identification with the 2MASS infrared catalogue (Cutri et al. 2003). The complete table is available in electronic form. 
\label{tab:detection}
}
\tablewidth{0pt}
\tablehead{
           \multicolumn{8}{c}{X-ray data}                                                                                                                                           & \multicolumn{5}{c}{2MASS NIR data}                                       &  \colhead{Mem}  &  \colhead{Mass}               &    \colhead{Remark}   \\
           \cline{9-13}
           \colhead{ID} & \colhead{$\rm{RA}$} &  \colhead{$\rm{DEC}$}& \colhead{err}               & \multicolumn{3}{c}{Count Rates}              &  \colhead{Dis}      & \colhead{N} & \colhead{off}      &  \colhead{$J$}&   \colhead{$H$}&   \colhead{$K_s$} &       \colhead{}  &   \colhead{}                 &       \colhead{}   \\
                        &                     &                      &                             & \multicolumn{3}{c}{$\rm{10^{-3}~cts~s^{-1}}$}&                     &             &       &  &  &    &         &                  &       \\
          \cline{5-7} 
          \colhead{}   &  \colhead{deg}      &   \colhead{deg}      & \colhead{\arcsec}            &    \colhead{PN}& \colhead{MOS1}&  \colhead{MOS2}         &  \colhead{\arcmin}  &              & \colhead{\arcsec}&   \colhead{mag}&  \colhead{mag}&    \colhead{mag}   &       \colhead{}   & \colhead{$\rm{M_\odot}$}       &      \colhead{} \\
           \colhead{1}&   \colhead{2}&           \colhead{3}&           \colhead{4}&                    \colhead{5}&   \colhead{6}&     \colhead{7}&            \colhead{8}         &  \colhead{9}&  \colhead{10}&        \colhead{11}&   \colhead{12}&     \colhead{13}      &     \colhead{14}   &      \colhead{15}           &  \colhead{16}        
}
\startdata
&&&&&&&&&&&&&&\\
\multicolumn{15}{c}{\bf{NGC 663}}\\
&&&&&&&&&&&&&&\\
\hline
&&&&&&&&&&&&&&\\
              1  &   26.177084  &   61.141083  &    1.3    &       0.0$\pm$0.0    &       1.8$\pm$0.5     &      1.5$\pm$0.5  &   13.7   &2   & 0.8    &     13.60$\pm$0.02     &    13.07$\pm$0.03    &     12.89$\pm$0.04    &     Y    &  2-1.4  &               \\           
              2  &   26.251165  &   61.158890  &    1.7    &       1.7$\pm$0.6    &       0.8$\pm$0.3     &      1.5$\pm$0.4  &   11.3   &1   & 2.1    &     15.08$\pm$0.04     &    14.58$\pm$0.06    &     14.34$\pm$0.08    &          &         &               \\
              3  &   26.279083  &   61.361000  &    0.7    &      10.7$\pm$1.3    &       2.8$\pm$0.6     &      2.5$\pm$0.6  &   13.1   &2   & 0.6    &     15.34$\pm$0.06     &    14.65$\pm$0.06    &     14.40$\pm$0.07    &          &         &               \\
              4  &   26.284582  &   61.317722  &    1.4    &       2.8$\pm$0.7    &       0.7$\pm$0.3     &      1.0$\pm$0.4  &   11.4   &2   & 0.8    &     14.07$\pm$0.03     &    13.46$\pm$0.04    &     13.29$\pm$0.04    &     Y    &  2-1.4  &               \\
\hline
\enddata
\begin{list}{}{}
\item[Notes -- ]
1:Identification number of X-ray sources detected using SAS task {\sc{edetect\_chain}}.\\ 
2: Right ascension of source in units of degree at J2000 epoch.\\
3: Declination of source in units of degree at J2000 epoch.\\
4: Error in the estimation of position from X-ray source detection algorithm in units of arcsec.\\
5: Count rates are estimated in energy band 0.3-7.5 keV from SAS task {\sc{edetect\_chain}} in PN detector.\\
6: Count rates are estimated in energy band 0.3-7.5 keV from SAS task {\sc{edetect\_chain}} in MOS1 detector.\\
7: Count rates are estimated in energy band 0.3-7.5 keV from SAS task {\sc{edetect\_chain}} in MOS2 detector.\\
8: Distances of the source from the center of the cluster in units of arcmin.\\
9 : Number of multiple identifications of a X-ray source in  2MASS NIR source catalogue within 10\arcsec search radius.\\
10: Distance (in arcsec) between the positions of the X-ray source and of its closest NIR within 10\arcsec search radius. 
    Only the closest identification of the X-ray source has been reported in this table.\\
11:  Magnitudes of the closest X-ray counterparts in  $J$ (1.25$\mu$m) band.\\
12:  Magnitudes of the  closest X-ray counterparts in  $H$ (1.65$\mu$m) band.\\
13:  Magnitudes of the  closest X-ray counterparts in  $K_s$ (2.17$\mu$m) band.\\
14: Membership of X-ray source in their corresponding cluster.'Y' represent the cluster member while 'N' represent non-member.\\
15: Masses of the sources estimated from color-magnitude diagrams of the clusters for cluster members in units of $\rm{M_\odot}$.\\
16: The information of the source from Vizier database. the references are : 
    Sk09 represent Skiff (2009);
    Cu10 represnt Currie et al. (2010);
    S02  represent Slesnick, Hillenbrand \& Massey (2002);
    S05  Strom, Wolff \& Dror (2005);
    Pic10 Pickles \& Depagne (2010);
    Ogu02 represent Ogura, Sugitani \& Andrew (2002);
    Ike08 represent Ikeda et al. (2008);
    Glebocki05 represent Glebocki \& Gnacinski (2005);
    D'orazi09 represent  D'Orazi \& Randich (2009);
    Delgado11 represent  Delgado, Alfaro \& Yun (2011); 
    Kher09 represnt Kharchenko \& Roeser (2009);
    Fab02 represent Fabricius et al. (2002).  
    The probability of the X-ray source for being a star is given from Flesch (2010; Fl10).\\ 
\end{list}
\end{deluxetable}
\end{landscape}

%% file: Cluater_Sources.tex
\begin{table*}
\caption{Detection fraction of X-ray sources within cluster radius with a comparison of radius of the clusters derived using NIR data and optical data.}
\label{tab:Source_info}      
\scriptsize
\begin{tabular}{ccccccccccc}  
\hline 
Cluster     &                           \multicolumn{2}{c}{Center}          & \multicolumn{2}{c}{Radius} & \multicolumn{4}{c} {No. Of X-ray Sources}                       &  Detection   \\ 
           \cline{2-3}    \cline{4-5}  \cline{6-9}  
Name        &                          \multicolumn{2}{c}{2MASS}            &   Dias02$^1$    & 2MASS     &  \multicolumn{2}{c}{Detected} & \multicolumn{2}{c}{Identified}  &  Limits  \\
            &                $\rm{RA_{J2000}}$   &  $\rm{DEC_{J2000}}$      &              &           &  Total    &  cluster          & Total & cluster                        &  $\rm {Log(L_X)}$        \\                    
            &                   hh:mm:ss         &  hh:mm:ss                &  ($^\prime$) &($^\prime$)&    &                   &       &                                & \egs      \\
\hline

NGC 663     &                 01:46:29           &  +61:13:05               &   7             &  15        &   85      &   84              &  60   &   32                     &  30.43      \\
NGC 869     &                 02:19:00           &  +57:08:57               &   9             &  14        &  183      &  181              & 141   &   78                     &  30.24       \\ 
NGC 884     &                 02:22:04           &  +57:08:49               &   9             &  11        &  147      &  114              & 103   &   71                     &  30.36    \\  
NGC 7380    &                 22:47:47           &  +58:07:15               &  10             &  8         &   88      &   40              &  76   &   37                     &  30.85   \\
Berkeley 86 &                 20:20:22           &  +38:42:07               &  3              &  3.5       &   96      &   11              &  90   &   11                     &  30.62   \\
IC 2602     &                 10:43:06           &  -64:25:30               &  50             &   --       &   95      &   95              &  74   &   74                     &  27.57      \\
Hogg 15     &                 12:43:39           &  -63:05:58               &  3.5            &  7         &  124      &   53              & 106   &   47                     &  30.70 \\ 
Trumpler 18 &                 11:11:31           &  -60:40:41               &  2.5            &   --       &  208      &   11              & 194   &   10                     &  29.75  \\  
\hline
\end{tabular}
\newline
1: radius derived using optical data in literature (Dias et al. 2002).\\
2: 2$\sigma$ detection limits of observations are derived from count rate conversion into flux in PN detector for low mass stars (see \S\ref{sec:lc_spec}).\\
\end{table*}

%% file: Massive_new1.tex
\begin{landscape}
\begin{table}
\caption{X-ray temperatures and luminosities of massive stars within young clusters.
  }
\label{tab:massive}      
\scriptsize
\begin{tabular}{c c c c c c c l c }
\hline
Cluster   & ID   & Name             & Spectral type        & $\rm{B^a}$ & $\rm{off^b}$  &   $\rm{kT_{av}}$      &       $\rm{Log(L_X)}$ (References)        &$\rm{Log(\frac{L_X}{L_{bol}})}$   \\ 

          &      &                  &                      &   & \arcsec  &  keV                      &         \egs                                  &                      \\
\hline
NGC663    & 26   & BD+60 329        &  B1V                 &   & 2.3  &                               &   31.16 (N$\acute{a}$ze 2009)                 &-6.69                  \\
NGC663    & 69   & V831 Cas         &  B1V                 & Y & 1.1  &  $\rm{1.11^{+0.07}_{-0.06}}$  &   34.15 (La Palombara \& Mereghetti 2006)     & --                     \\
NGC869    & 12   & HD 13969         &  B0.5I               &   & 1.8  &                               &   30.53$^{\dagger\dagger}$ (Present study)    &-7.37                   \\ 
NGC869    & 50   & HD 14052         &  B1.5I               &   & 4.3  &  $\rm{0.44^{+0.18}_{-0.11}}$  &   31.06 (Present study)                       &-7.34                   \\ 
NGC869    &103   & [SHM2002] 120    &  B1.5V               &   & 6.3  &                               &   30.84$^{\dagger\dagger}$ (Present study)    &-7.18                   \\ 
NGC869    &109   & BD+56 527        &  B2I                 &   & 2.0  &  $\rm{1.12^{+0.37}_{-0.21}}$  &   31.02 (Present study)                       &-7.33                   \\  
NGC869    &119   & [SHM2002] 138    &  B                   &   & 5.6  &  $\rm{4.64^{+21.04}_{-2.5}}$  &   31.09 (Present study)                       &-6.43                    \\
NGC884    & 79   & [SHM2002] 131    &  B1.5III             &   & 4.4  &                               &   30.80 (N$\acute{a}$ze 2009)                 &-6.76  \\
NGC884    & 93   & BD+56 578        &  B2III               &   & 2.2  &                               &   30.89 (N$\acute{a}$ze 2009)                 &-7.14  \\
NGC7380   & 36   & DH Cep           &  O5.5V+O6.5V         & Y & 0.6  &  $\rm{0.64^{+0.02}_{-0.02}}$  &   32.43 (Bhatt et al. 2010)                   &-6.71  \\
NGC7380   & 77   & LS III +57 90    &  O8V((f))            & Y & 1.7  &                               &   31.55 (N$\acute{a}$ze 2009)                 &-6.66  \\
Be86      & 85   & HD 228989        &  O9V+O9V             & Y & 2.7  &  $\rm{0.62^{+0.15}_{-0.11}}$  &   31.72  (Present study)                      &-6.62 \\
Hogg15    &  3   & HD 110432        &  B0.5Vep             & Y & 2.2  &  8.0 --11.0                   &   35.00 (Lopes de Oliveira et al. 2007)       &--     \\
Hogg15    & 49   & MO 1- 78         &  OB                  &   & 1.9  &  $\rm{0.72^{+0.23}_{-0.17}}$  &   31.39  (Present study)                      &-6.91  \\
Hogg15    & 69   &                  &                      &   & 2.1  &                               &   30.94$^{\dagger\dagger}$   (Present study)  &-6.72  \\
Hogg15    & 73   &                  &                      &   & 0.4  &                               &   30.94$^{\dagger\dagger}$   (Present study)  &-7.02  \\
\hline
\end{tabular}
\newline
a: 'Y' represents the binarity of stars from literature.\\ 
b: offset between the position of Massive stars in 2MASS catalogue and their X-ray counterparts.\\
$\dagger\dagger$ :  $\rm{L_X}$ has been derived from the X-ray flux estimated from the count rate conversion in PN detector using the WebPIMMS,i.e.,  $\rm{Flux= CCF  \times Count rate}$
The values of CCFs (in units of $\rm{erg~s^{-1}~cm^{2}}$)  have been derived for massive stars in PN  are  $\rm{3.606\times10^{-12}}$ for
NGC 869 and  $\rm{7.689\times10^{-12}}$  for Hogg 15, respectively.
\end{table}
\end{landscape}

%% file: spectral_properties_new1.tex
\begin{deluxetable}{cccccccccc}
\tabletypesize{\tiny}
\tablecaption{\scriptsize Spectral and Timing properties of Intermediate and Low mass stars in young open clusters.
\label{tab:Int_lo_spec}
}
\tablewidth{-0pt}
\tablehead{
         \multicolumn{4}{c}{Source Detection} &
         \multicolumn{4}{c}{Spectral } &
         \multicolumn{2}{c}{Timing }\\ 
       \colhead{Cluster} &
        \colhead{ID} & 
        \colhead{$\rm{RA_{J2000}}$} &
       \colhead{$\rm{DEC_{J2000}}$} &
       \colhead{kT} &
       \colhead{Log(EM)} &
       \colhead{$\rm{Log(L_X)^{\dagger}}$ }  &
       \colhead{$\rm{Log(\frac{L_X}{L_{bol}})}$ } &
       \colhead{Bin Size } &
       \colhead{$\rm{F_{var}}$ }  \\
       \colhead{Name} &
        \colhead{} & 
        \colhead{deg} &
       \colhead{deg} &
       \colhead{keV} &
       \colhead{$\rm{cm^{-3}}$} &
       \colhead{$\rm {ergs~s^{-1}}$} &
       \colhead{} &
       \colhead{s} & 
       \colhead{} 
       }
\startdata
\hline 
                                           \multicolumn{10}{c}{Intermediate mass stars ($\rm{2--10 M_\odot}$)}  \\ 
\hline    
    NGC663      &      42   &   26.583460  &    61.263973     &       $>2.78$                      &    $53.81^{+0.14}_{-0.18}$      & 31.02   &      -6.47   &       &                  \\
    NGC663      &      48   &   26.604250  &    61.208221     &       $>1.79$                      &    $53.85^{+0.15}_{-0.19}$      & 30.98   &      -6.01   &       &                   \\
    NGC869      &      10   &   34.441750  &    57.088139     &                                    &                                 & 30.70   &      -5.75   &       &                   \\                                   
    NGC869      &      13   &   34.457584  &    57.318527     &                                    &                                 & 31.26   &      -4.75   &       &                   \\      
    NGC869      &      17   &   34.481415  &    57.218498     &                                    &                                 & 30.98   &      -6.08   &       &                   \\          
    NGC869      &      21   &   34.512001  &    57.076889     &                                    &                                 & 30.51   &      -4.43   &       &                   \\          
    NGC869      &      64   &   34.647793  &    57.214417     &                                    &                                 & 30.81   &      -4.91   &       &                   \\             
    NGC869      &      67   &   34.662083  &    57.220444     & $0.54^{+0.43}_{-0.33}$             &    $53.42^{+0.56}_{-0.32}$      & 30.48   &      -4.35   &800    & 0.89$\pm$0.17     \\       
    NGC869      &      69   &   34.671417  &    57.102196     & $2.75^{+14.10}_{-1.45}$            &    $54.03^{+0.16}_{-0.21}$      & 31.10   &      -3.50   &       &                    \\
    NGC869      &      88   &   34.750916  &    57.154499     &                                    &                                 & 30.51   &      -6.47   &       &                    \\          
    NGC869      &     105   &   34.783291  &    57.127804     &  $3.10^{+2.50}_{-1.10}$            &    $54.01^{0.10}_{-0.10}$       & 31.11   &      -4.55   &       &                    \\       
    NGC869      &     139   &   34.869793  &    57.154026     &  $1.95^{+1.96}_{-0.62}$            &    $54.06^{+0.09}_{-0.11}$      & 31.09   &      -6.16   &       &                    \\       
    NGC869      &     140   &   34.870708  &    57.163887     &  $1.27^{+0.43}_{-0.23}$            &    $53.98^{+0.13}_{-0.15}$      & 31.04   &      -5.42   &800    & 0.72$\pm$0.14      \\       
    NGC869      &     169   &   35.026749  &    57.125500     &                                    &                                 & 30.70   &      -6.53   &       &                     \\  
    NGC869      &     173   &   35.037834  &    57.020638     &  $2.80^{+2.30}_{-0.80}$            &    $54.34^{0.08}_{-0.10}$       & 31.52   &      -5.04   &       &                     \\       
    NGC884      &      23   &   35.328167  &    57.146667     &       $>3.80$                      &    $54.12^{+0.20}_{-0.09}$      & 31.33   &      -4.84   &1100   & 0.13$\pm$0.47       \\
    NGC884      &      62   &   35.461918  &    57.026669     &       $>3.26$                      &    $53.90^{+0.22}_{-0.17}$      & 31.10   &      -5.08   &       &                      \\
    NGC884      &      68   &   35.489666  &    57.211613     &                                    &                                 &$<$30.52 &   $<$-6.22   &       &                     \\
    NGC884      &      78   &   35.517044  &    57.142250     &                                    &                                 & 30.81   &      -5.12   &       &                     \\
    NGC884      &     103   &   35.613209  &    57.054779     &       $1.30^{+0.31}_{-0.23}$       &    $54.00^{+0.11}_{-0.14}$      & 31.04   &      -6.29   &       &                     \\
    NGC884      &     120   &   35.707790  &    57.142834     &       $>15.80$                     &    $54.06^{+0.13}_{-0.24}$      & 31.16   &      -4.18   &       &                     \\
   NGC7380      &      51   &   341.83798  &    58.088665     &                                    &                                 & 30.99   &      -5.44   &       &                     \\
   NGC7380      &      59   &   341.89688  &    58.125500     &       $2.38^{+1.43}_{-0.65}$       &    $54.59^{+0.08}_{-0.09}$      & 31.65   &      -4.45   &       &                     \\
  Berkeley 86   &      76   &   305.04416  &    38.695137     &      $1.36^{+0.70}_{-0.31}$        &    $53.74^{+0.17}_{-0.20}$      & 30.47   &      -5.80   &       &                     \\       
  Berkeley 86   &      79   &   305.00755  &    38.680832     &       \multicolumn{4}{c}{Not fitted with model}                                               &       &                     \\
    Hogg15      &      11   &   190.79124  &   -63.003056     &       $0.14^{+0.62}_{-0.05}$       &    $>56.95$                     & 32.57   &      -4.44   &       &                     \\
    Hogg15      &      33   &   190.89438  &   -63.014778     &       $0.28^{+0.66}_{-0.10}$       &    $>55.45$                     & 31.73   &      -4.86   &       &                      \\
    Hogg15      &      34   &   190.98633  &   -63.081444     &   $ {1.55^{+0.27}_{-0.24}}$        &    $54.80^{+0.06}_{-0.06}$      & 31.76   &      -5.46   &       &                      \\
    Hogg15      &      42   &   190.96179  &   -63.113388     &       $0.16^{+0.12}_{-0.05}$       &    $>56.17$                     & 32.13   &      -4.09   &       &                     \\
    Hogg15      &      48   &   190.98618  &   -63.071609     &       $0.18^{+0.05}_{-0.03}$       &    $55.44^{+0.38}_{-0.36}$      & 32.19   &      -4.15   &       &                     \\
    Hogg15      &      55   &   191.00912  &   -63.067471     &                                    &                                 & 31.71   &      -5.01   &       &                     \\
    Hogg15      &      65   &   191.06125  &   -63.137749     &       $0.69^{+0.26}_{-0.22}$       &    $54.20^{+0.16}_{-0.17}$      & 31.30   &      -4.85   &4800   & 0.69$\pm$0.33       \\
    Hogg15      &      82   &   191.15871  &   -63.086193     &       $>6.27$                      &    $54.39^{+0.12}_{-0.19}$      & 31.56   &      -4.87   &       &                     \\
  Trumpler 18   &      76   &   167.88104  &   -60.664612     &       $6.99^{+50.8}_{-3.61}$       &    $53.53^{+0.08}_{-0.09}$      & 30.70   &      -4.50   &       &                     \\
  Trumpler 18   &      80   &   167.89404  &   -60.667278     &       $5.63^{+13.8}_{-2.82}$       &    $54.21^{+0.08}_{-0.08}$      & 31.38   &      -4.77   &       &                     \\
  Trumpler 18   &      90   &   167.92091  &   -60.706276     &       $0.84^{+0.12}_{-0.08}$       &    $53.71^{+0.05}_{-0.06}$      & 30.81   &      -5.75   &       &                     \\
\hline                                                                                                                                                
    \multicolumn{10}{c}{Low mass stars ($\rm{<2 M_\odot}$)}\\                                                                                         
\hline                                                                                                                                                
    NGC663     &        1   &  26.177084  &   61.141083    &                                   &                                &       31.23      &       -4.17   &           &                  \\        
    NGC663     &        4   &  26.284582  &   61.317722    &                                   &                                &       30.98      &       -4.02   &           &                  \\       
    NGC663     &        7   &  26.319500  &   61.133999    &                                   &                                &       30.79      &       -4.28   &           &                  \\        
    NGC663     &        8   &  26.320749  &   61.281723    &                                   &                                &       31.33      &       -4.15   &           &                  \\           
    NGC663     &       12   &  26.358292  &   61.181416    &                                   &                                &       30.98      &       -4.58   &           &                  \\       
    NGC663     &       15   &  26.378542  &   61.141777    &                                   &                                &       31.36      &       -3.57   &           &                  \\       
    NGC663     &       17   &  26.385792  &   61.345470    &                                   &                                &       30.96      &       -3.68   &           &                  \\       
    NGC663     &       18   &  26.390625  &   61.064499    &                                   &                                &       30.95      &       -3.59   &           &                  \\       
    NGC663     &       41   &  26.582500  &   61.229389    &                                   &                                &       30.93      &       -4.37   &           &                  \\       
    NGC663     &       52   &  26.639708  &   61.056168    &                                   &                                &       30.85      &       -3.41   &           &                  \\       
    NGC663     &       56   &  26.663834  &   61.150028    &       $0.27^{+0.02}_{-0.02}$      &       $55.29^{+0.09}_{-0.07}$  &       32.16(?)   &       -2.01(?)& 500       &  0.10$\pm$0.23    \\      
    NGC663     &       61   &  26.689041  &   61.107445    &       $>9.66$                     &       $54.35^{+0.08}_{-0.18}$  &       31.46(?)   &       -1.84(?)&           &                  \\      
    NGC663     &       66   &  26.729250  &   61.048973    &                                   &                                &       31.14      &       -3.42   &           &                   \\      
    NGC663     &       80   &  26.941376  &   61.183498    &       $>9.37$                     &       $54.64^{+0.11}_{-0.12}$  &       31.78      &       -3.69   & 2000      &  0.26$\pm$0.18    \\      
    NGC663     &       81   &  26.983583  &   61.209415    &                                   &                                &       31.57      &       -3.24   &           &                   \\      
    NGC663     &       82   &  27.015417  &   61.212749    &       $1.36^{+0.79}_{-0.38}$      &       $54.15^{+0.16}_{-0.19}$  &       31.17      &       -4.17   &           &                    \\      
    NGC663     &       83   &  27.019709  &   61.118973    &       $0.50^{+0.29}_{-0.31}$      &       $54.74^{+0.79}_{-0.26}$  &       31.76      &       -3.52   & 3500      &  0.54$\pm$0.40    \\      
    NGC869     &       1    &  34.341042  &   57.174751    &                                   &                                &       31.16      &       -2.96   &           &                            \\ 
    NGC869     &       19   &  34.498043  &   56.965668    &                                   &                                &       31.01      &       -3.85   &           &                            \\
    NGC869     &       20   &  34.511124  &   57.110722    &                                   &                                &       30.54      &       -3.51   &           &                            \\
    NGC869     &       22   &  34.518623  &   57.345470    &                                   &                                &       30.91      &       -3.92   &           &                            \\
    NGC869     &       23   &  34.522251  &   57.024361    &                                   &                                &       30.65      &       -3.40   &           &                            \\  
    NGC869     &       25   &  34.530666  &   57.207554    &                                   &                                &       30.57      &       -3.96   &           &                            \\
    NGC869     &       33   &  34.557877  &   57.297722    &                                   &                                &       30.65      &       -4.03   &           &                            \\
    NGC869     &       34   &  34.559208  &   57.030613    &                                   &                                &       30.78      &       -4.18   &           &                            \\  
    NGC869     &       36   &  34.571877  &   57.125416    &          $0.97^{+0.65}_{-0.19}$   &       $53.85^{+0.12}_{-0.14}$  &       30.95      &       -3.16   &           &                            \\      
    NGC869     &       40   &  34.584126  &   57.273140    &                                   &                                &       30.60      &       -3.36   &           &                            \\
    NGC869     &       41   &  34.585751  &   57.109943    &                                   &                                &       31.03      &       -3.46   &           &                            \\
    NGC869     &       42   &  34.586708  &   57.174110    &          $3.66^{+14.00}_{-1.90}$  &       $54.25^{+0.11}_{-0.12}$  &       31.38      &       -3.18   &600        &  0.47$\pm$0.10            \\      
    NGC869     &       44   &  34.595165  &   57.182278    &          $2.47^{+3.82}_{-0.85}$   &       $53.86^{+0.12}_{-0.14}$  &       30.93      &       -3.36   &           &                           \\      
    NGC869     &       47   &  34.603626  &   57.112083    &                                   &                                &       30.47      &       -3.87   &           &                           \\      
    NGC869     &       58   &  34.628918  &   57.064999    &          $>2.21$                  &       $53.65^{+0.28}_{-0.24}$  &       30.86      &       -3.05   &           &                            \\      
    NGC869     &       61   &  34.633835  &   57.248390    &                                   &                                &       30.65      &       -4.51   &           &                            \\
    NGC869     &       68   &  34.668709  &   57.127224    &                                   &                                &       30.67      &       -3.59   &           &                            \\
    NGC869     &       73   &  34.696793  &   56.956944    &                                   &                                &       31.09      &       -3.08   &           &                            \\
    NGC869     &       74   &  34.701958  &   57.074444    &          $1.25^{+0.77}_{-0.27}$   &       $53.75^{+0.15}_{-0.16}$  &       30.80      &       -4.53   &3000       &  0.20$\pm$0.55             \\      
    NGC869     &       75   &  34.703083  &   57.121082    &          $2.60^{+1.90}_{-0.75}$   &       $54.11^{+0.08}_{-0.09}$  &       31.19      &       -3.85   &           &                   \\      
    NGC869     &       78   &  34.717876  &   57.192001    &          $3.30^{+7.45}_{-1.38}$   &       $53.84^{+0.13}_{-0.15}$  &       29.69      &       -4.46   &           &                           \\      
    NGC869     &       79   &  34.719166  &   57.074360    &                                   &                                &       30.44      &       -3.86   &           &                   \\      
    NGC869     &       92   &  34.754833  &   57.003334    &          $>2.28$                  &       $54.04^{+0.12}_{-0.13}$  &       31.24      &       -3.82   &           &                   \\      
    NGC869     &       93   &  34.754959  &   57.097694    &                                   &                                &       30.85      &       -3.83   &           &                  \\       
    NGC869     &       96   &  34.759666  &   57.160721    &          $3.00^{+3.94}_{-1.30}$   &       $54.07^{+0.10}_{-0.12}$  &       31.17      &       -3.82   &2000       &  0.31$\pm$0.27    \\      
    NGC869     &       99   &  34.768124  &   57.160999    &          $2.42^{+2.27}_{-0.90}$   &       $54.07^{+0.09}_{-0.10}$  &       31.13      &       -3.27   &1000       &  0.45$\pm$0.21    \\      
    NGC869     &       104  &  34.782333  &   57.102638    &          $>1.89$                  &       $53.83^{+0.13}_{-0.16}$  &       30.98      &       -3.23   &           &                            \\
    NGC869     &       108  &  34.790833  &   57.153000    &                                   &                                &       30.35      &       -4.18   &           &                   \\      
    NGC869     &       111  &  34.794624  &   57.125721    &                                   &                                &                  &               &600        &   0.64$\pm$0.09                 \\      
    NGC869     &       113  &  34.801960  &   57.064667    &          $0.23^{+0.05}_{-0.04}$   &       $54.40^{+0.21}_{-0.17}$  &       31.25      &       -3.67   &           &                            \\
    NGC869     &       114  &  34.807625  &   57.030193    &          $1.71^{+1.25}_{-0.520}$  &       $53.79^{+0.14}_{-0.20}$  &       30.80      &       -3.24   &           &                            \\
    NGC869     &       115  &  34.812290  &   57.218582    &                                   &                                &       30.80      &       -3.22   &           &                            \\
    NGC869     &       121  &  34.824207  &   57.134972    &                                   &                                &       30.35      &       -4.03   &           &                            \\
    NGC869     &       123  &  34.830917  &   57.202499    &                                   &                                &       30.68      &       -3.93   &           &                            \\
    NGC869     &       128  &  34.839500  &   57.034805    &                                   &                                &       30.90      &       -3.34   &           &                            \\
    NGC869     &       129  &  34.841667  &   57.232334    &                                   &                                &       30.57      &       -3.30   &           &                            \\
    NGC869     &       132  &  34.848293  &   57.025417    &                                   &                                &       30.72      &       -3.79   &           &                            \\
    NGC869     &       135  &  34.854874  &   56.991890    & $0.99^{+0.73}_{-0.64}$            &       $54.07^{+0.15}_{-0.17}$  &       31.17      &       -3.60   &           &                            \\
    NGC869     &       143  &  34.883999  &   57.113998    &                                   &                                &       30.54      &       -4.32   &           &                            \\
    NGC869     &       145  &  34.890583  &   57.059082    & $0.73^{+0.28}_{-0.27}$            &  $53.60^{+0.15}_{-0.19}$       &       30.72      &       -3.24   &           &                            \\
    NGC869     &       146  &  34.890751  &   57.258167    &                                   &                                &       30.51      &       -4.10   &           &                            \\
    NGC869     &       149  &  34.91346   &   57.060806    &                                   &                                &       30.51      &       -4.22   &           &                            \\
    NGC869     &       152  &  34.929249  &   57.312527    &                                   &                                &       30.83      &       -3.16   &           &                            \\
    NGC869     &       155  &  34.947918  &   57.151196    &                                   &                                &       31.20      &       -3.38   &           &                            \\
    NGC869     &       156  &  34.952667  &   57.039165    & $4.40^{+12.0}_{-1.90}$            &  $54.12^{+0.10}_{-0.12}$       &       31.28      &       -3.21   &           &                            \\
    NGC869     &       158  &  34.960876  &   57.251278    &                                   &                                &       30.81      &       -3.64   &           &                            \\
    NGC869     &       159  &  34.961082  &   57.199333    &          $2.14^{+2.54}_{-0.69}$   &       $54.09^{+0.10}_{-0.12}$  &       31.12      &       -3.31   &           &                            \\
    NGC869     &       162  &  34.980000  &   57.220444    &          $0.93^{+0.30}_{-0.47}$   &       $53.80^{+0.14}_{-0.17}$  &       30.96      &       -3.85   &           &                            \\
    NGC869     &       165  &  34.999458  &   57.253887    &                                   &                                &       30.74      &       -4.01   &           &                            \\
    NGC869     &       166  &  35.011665  &   57.318638    &                                   &                                &       31.15      &       -2.52   &           &                            \\
    NGC869     &       174  &  35.071041  &   57.147530    &                                   &                                &       30.93      &       -2.95   &           &                            \\
    NGC869     &       175  &  35.083668  &   57.010471    &                                   &                                &    $<$30.65      &    $<$-3.78   &           &                            \\
    NGC884     &       16   &  35.268585  &   57.129139    &                                   &                                &       30.56      &       -4.26   &           &                            \\
    NGC884     &       28   &  35.357166  &   57.050220    &                                   &                                &       30.59      &       -4.16   &           &                            \\
    NGC884     &       30   &  35.369831  &   57.103806    &       $1.06^{+0.42}_{-0.35}$      &       $53.77^{+0.44}_{-0.35}$  &       30.75      &       -3.71   &           &                           \\
    NGC884     &       39   &  35.409168  &   57.068306    &                                   &                                &       30.47      &       -3.76   &           &                          \\
    NGC884     &       49   &  35.434875  &   57.264473    &                                   &                                &       30.74      &       -3.53   &           &                          \\
    NGC884     &       53   &  35.442039  &   57.044498    &       $>0.70$                     &       $<54.45$                 &       30.94      &       -3.37   &           &           \\
    NGC884     &       64   &  35.466541  &   57.276859    &       $2.40^{+5.20}_{-0.94}$      &       $54.00^{+0.14}_{-0.17}$  &       31.07      &       -2.84   &           &           \\
    NGC884     &       71   &  35.503956  &   57.079613    &       $>4.26$                     &       $54.06^{+0.17}_{-0.09}$  &       31.27      &       -3.67   &           &           \\
    NGC884     &       75   &  35.512707  &   57.308529    &                                   &                                &       31.09      &       -3.13   &           &                            \\
    NGC884     &       80   &  35.524750  &   57.118694    &                                   &                                &       30.50      &       -4.39   &           &                              \\
    NGC884     &       88   &  35.552002  &   57.100750    &       $1.65^{+0.68}_{-0.31}$      &       $54.15^{+0.09}_{-0.10}$  &       31.17      &       -2.81   & 2200      &   0.43$\pm$ 0.17           \\
    NGC884     &       89   &  35.553165  &   57.176613    &                                   &                                &       30.98      &       -3.29   &           &                           \\
    NGC884     &       90   &  35.563168  &   57.140194    &                                   &                                &       30.79      &       -3.21   &           &                      \\
    NGC884     &       91   &  35.567169  &   57.094028    &       $1.83^{+1.56}_{-0.47}$      &       $54.01^{+0.10}_{-0.11}$  &       31.04      &       -3.04   & 2500      &   0.12$\pm$ 0.62   \\      
    NGC884     &      101   &  35.605461  &   57.012085    &                                   &                                &       30.92      &       -3.20   &           &                     \\    
    NGC884     &      102   &  35.611462  &   57.089748    &                                   &                                &       30.64      &       -3.87   &           &                     \\    
    NGC884     &      104   &  35.614544  &   57.025196    &                                   &                                &       30.66      &       -4.33   &           &                     \\      
    NGC884     &      106   &  35.626041  &   57.242638    &                                   &                                &       30.59      &       -3.41   &           &                     \\      
    NGC884     &      110   &  35.643375  &   57.191418    &                                   &                                &       30.47      &       -4.65   &           &                     \\      
    NGC884     &      113   &  35.655708  &   57.217945    &                                   &                                &       30.61      &       -3.35   &           &                     \\      
    NGC884     &      119   &  35.707542  &   57.161026    &                                   &                                &       31.11      &       -3.42   &           &                     \\      
    NGC884     &      121   &  35.722958  &   57.098804    &                                   &                                &    $<$30.53      &    $<$-4.41   &           &                     \\     
    NGC884     &      122   &  35.724209  &   57.126083    &       $1.02^{+0.27}_{-0.21}$      &       $53.88^{+0.12}_{-0.14}$  &       30.96      &       -3.78   &           &                      \\
    NGC884     &      130   &  35.747002  &   57.137417    &                                   &                                &       30.44      &       -4.08   &           &                            \\
    NGC884     &      137   &  35.803707  &   57.122444    &                                   &                                &       30.74      &       -3.42   &           &                            \\
    NGC884     &      138   &  35.820457  &   57.222500    &                                   &                                &       31.27      &       -3.29   &           &                            \\
   NGC7380     &       34   &  341.70270  &   58.133194    &                                   &                                &       30.86      &       -4.59   &           &                           \\
   NGC7380     &       35   &  341.72348  &   58.168083    &                                   &                                &    $<$31.28      &    $<$-2.87   &           &                           \\
   NGC7380     &       37   &  341.72653  &   58.134724    &                                   &                                &       30.88      &       -3.41   &           &                            \\
   NGC7380     &       41   &  341.74469  &   58.076694    &                                   &                                &       30.92      &       -3.16   &           &                            \\
   NGC7380     &       46   &  341.77658  &   58.113888    &                                   &                                &       31.09      &       -2.81   &           &                            \\
   NGC7380     &       48   &  341.80121  &   58.109417    &       $1.78^{+2.35}_{-0.60}$      &       $54.06^{+0.16}_{-0.22}$  &       31.09      &       -3.30   &           &                        \\
   NGC7380     &       49   &  341.80286  &   58.191334    &                                   &                                &       31.06      &       -3.29   &           &                            \\
   NGC7380     &       50   &  341.81659  &   58.067890    &       $1.37^{+0.51}_{-0.22}$      &       $54.18^{+0.13}_{-0.15}$  &       31.20      &       -3.20   &           &                         \\
   NGC7380     &       57   &  341.89398  &   58.138054    &                                   &                                &       31.25      &       -3.27   &           &                            \\
   NGC7380     &       60   &  341.89981  &   58.049862    &                                   &                                &       31.08      &       -2.72   &           &                            \\
   NGC7380     &       62   &  341.90005  &   58.075222    &       $2.72^{+3.98}_{-0.92}$      &       $54.36^{+0.12}_{-0.14}$  &       31.45      &       -3.12   &           &                        \\
   NGC7380     &       65   &  341.91592  &   58.099998    &                                   &                                &       30.96      &       -3.00   &           &                              \\
   NGC7380     &       66   &  341.91595  &   58.023472    &       $>4.68$                     &       $54.48^{+0.18}_{-0.14}$  &       31.69      &       -2.75   &1200       &  0.15$\pm$0.53             \\
   NGC7380     &       67   &  341.92245  &   58.148861    &       $1.64^{+1.25}_{-0.46}$      &       $54.44^{+0.12}_{-0.14}$  &       31.48(?)   &       -2.34(?)&           &                     \\
   NGC7380     &       70   &  341.93970  &   58.058472    &                                   &                                &       31.13      &       -2.99   &           &                            \\
   NGC7380     &       71   &  341.94296  &   58.184776    &                                   &                                &       31.12(?)   &       -2.49(?)&           &                      \\
   NGC7380     &       80   &  341.97977  &   58.056499    &                                   &                                &       31.14      &       -2.69   &           &                      \\
   NGC7380     &       82   &  342.00183  &   58.044971    &                                   &                                &       31.03      &       -3.20   &           &                      \\
   NGC7380     &       83   &  342.01785  &   58.068554    &                                   &                                &       31.22      &       -2.85   &           &                            \\
   NGC7380     &       85   &  342.02817  &   58.075195    &       $6.50^{+4.31}_{-2.22}$      &       $55.03^{+0.05}_{-0.05}$  &       32.22      &       -2.06   &           &                        \\
   NGC7380     &       86   &  342.12955  &   58.147583    &                                   &                                &       31.43      &       -4.00   &           &                             \\
  Berkeley 86  &       75   &  305.04169  &   38.721001    &       $1.87^{+1.63}_{-0.63}$      &       $54.15^{+0.10}_{-0.12}$  &       31.18      &       -4.24   &2500       &    0.22$\pm$0.34          \\
  Berkeley 86  &       81   &  305.08072  &   38.700722    &                                   &                                &       30.92(?)   &       -2.27(?)&           &                             \\
  Berkeley 86  &       84   &  305.08899  &   38.687695    &                                   &                                &    $<$30.78      &    $<$-4.51   &           &                             \\
  Berkeley 86  &       89   &  305.10193  &   38.700638    &                                   &                                &       31.21      &       -4.50   &           &                              \\
  Berkeley 86  &       93   &  305.12442  &   38.665749    &                                   &                                &       30.98      &       -2.75   &           &                                \\
    IC2602     &        6   &  160.24988  &  -64.334000    &       $1.31^{+0.31}_{-0.09}$      &       $52.21^{+0.05}_{-0.06}$  &       29.24      &       -3.43   &400        &   1.02$\pm$0.09     \\    
    IC2602     &       16   &  160.36188  &  -64.339302    &       $2.33^{+4.80}_{-0.74}$      &       $51.34^{+0.13}_{-0.16}$  &       27.80      &       -3.67   &           &                   \\    
    IC2602     &       20   &  160.43851  &  -64.467941    &       $1.22^{+0.05}_{-0.05}$      &       $52.64^{+0.03}_{-0.03}$  &       29.68      &       -3.18   &300        &   0.19$\pm$0.05     \\    
    IC2602     &       48   &  160.67279  &  -64.351387    &       $0.94^{+0.01}_{-0.01}$      &       $53.22^{+0.01}_{-0.01}$  &       30.29      &       -3.41   &400        &   0.25$\pm$0.02     \\    
    IC2602     &       54   &  160.71042  &  -64.364891    &       $0.93^{+0.09}_{-0.20}$      &       $51.55^{+0.06}_{-0.07}$  &       28.65      &       -3.06   &1500       &   0.22$\pm$0.16     \\    
    IC2602     &       66   &  160.81512  &  -64.398415    &       $0.62^{+0.04}_{-0.04}$      &       $52.13^{+0.03}_{-0.03}$  &       29.22      &       -4.37   &800        &   0.13$\pm$0.08     \\    
    IC2602     &       67   &  160.84492  &  -64.486832    &       $0.80^{+0.19}_{-0.08}$      &       $51.88^{+0.05}_{-0.05}$  &       28.99      &       -3.21   &1000       &   0.21$\pm$0.14     \\    
    IC2602     &       87   &  161.04083  &  -64.247528    &       $0.95^{+0.06}_{-0.18}$      &       $52.12^{+0.05}_{-0.05}$  &       29.22      &       -3.39   &1000       &   0.19$\pm$0.10     \\    
    IC2602     &       90   &  161.08730  &  -64.502136    &       $0.72^{+0.24}_{-0.29}$      &       $51.70^{+0.12}_{-0.14}$  &       28.81      &       -3.08   &           &                      \\    
    IC2602     &       91   &  161.09322  &  -64.258446    &       $0.73^{+0.07}_{-0.07}$      &       $52.00^{+0.05}_{-0.05}$  &       29.11      &       -4.40   &1000       &    0.28$\pm$0.12     \\    
    Hogg15     &       10   &  190.79047  &  -63.076221    &                                   &                                &       31.03      &       -3.43   &           &                   \\                                
    Hogg15     &       13   &  190.80016  &  -63.146889    &                                   &                                &       31.15      &       -3.34   &           &                   \\        
    Hogg15     &       14   &  190.80179  &  -63.102390    &       $0.96^{+0.31}_{-0.27}$      &       $54.21^{+0.14}_{-0.16}$  &       31.31      &       -4.69   &           &                     \\        
    Hogg15     &       16   &  190.82042  &  -63.084251    &       $0.63^{+0.19}_{-0.40}$      &       $54.35^{+0.72}_{-0.17}$  &       31.45      &       -3.96   &           &                     \\        
    Hogg15     &       18   &  190.82851  &  -63.083611    &                                   &                                &       31.06      &       -4.98   &           &                   \\        
    Hogg15     &       21   &  190.83771  &  -63.015751    &       $0.830^{+0.14}_{-0.21}$     &       $54.44^{+0.13}_{-0.13}$  &       31.55      &       -4.18   &           &                     \\        
    Hogg15     &       24   &  190.85812  &  -63.105526    &                                   &                                &       31.56      &       -2.67   &           &                   \\        
    Hogg15     &       30   &  190.89046  &  -63.065945    &                                   &                                &       31.03      &       -4.70   &           &                   \\        
    Hogg15     &       31   &  190.89145  &  -63.073444    &                                   &                                &       30.92      &       -5.40   &           &                   \\        
    Hogg15     &       32   &  190.89250  &  -63.001778    &                                   &                                &       30.92      &       -4.66   &           &                   \\        
    Hogg15     &       35   &  190.90747  &  -63.157776    &       $0.510^{+0.24}_{-0.21}$     &       $54.49^{+0.36}_{-0.20}$  &       31.54      &       -3.83   &           &        \\        
    Hogg15     &       37   &  190.93520  &  -63.099888    &       $2.34^{+0.81}_{-0.49}$      &       $54.71^{+0.06}_{-0.07}$  &       31.77      &       -2.59   &1000       &     0.30$\pm$0.15   \\        
    Hogg15     &       50   &  190.98984  &  -63.164001    &       $0.640^{+0.19}_{-0.13}$     &       $54.64^{+0.11}_{-0.13}$  &       31.73      &       -4.04   &           &                     \\        
    Hogg15     &       52   &  190.99983  &  -63.008915    &                                   &                                &       30.76      &       -3.49   &           &                   \\        
    Hogg15     &       56   &  191.00992  &  -63.171501    &                                   &                                &       30.92      &       -3.41   &           &                   \\        
    Hogg15     &       70   &  191.09630  &  -63.062611    &                                   &                                &       31.42      &       -4.18   &           &                   \\        
    Hogg15     &       71   &  191.09637  &  -63.123444    &       $0.860^{+0.21}_{-0.27}$     &       $54.12^{+0.13}_{-0.16}$  &       31.33      &       -4.11   &           &                     \\        
    Hogg15     &       72   &  191.10333  &  -63.099804    &                                   &                                &       30.87      &       -4.68   &           &                   \\        
  Trumpler 18  &       79   &  167.89359  &  -60.655224    &       $1.92^{+2.87}_{-0.75}$      &       $53.36^{+0.10}_{-0.12}$  &       30.38      &       -5.26   &           &                     \\    
  Trumpler 18  &       84   &  167.90800  &  -60.666779    &       $>26.50$                    &       $53.82^{+0.07}_{-0.11}$  &       30.92      &       -2.75   &           &                     \\    
  Trumpler 18  &       85   &  167.90945  &  -60.679501    &       $12.60^{+30.50}_{-5.97}$    &       $53.86^{+0.10}_{-0.06}$  &       31.05      &       -3.54   &           &                     \\    
\hline                                                                                                                                                                             
\enddata
\begin{list}{}{}
\item[Notes -- ]
{\tiny {
 Column 1 : Cluster name; 
 Column 2 : identification number (ID) in Table~\ref{tab:detection};
 Column 3 and  4 represent the position of X-ray source; 
 Column 5, 6 ,7 : estimated values of coronal temperatures (kT), emission measure (EM) and X-ray luminosities ($\rm{Log(L_X)}$) from either spectral fitting using C-statistics or
derived from conversion of count rates into X-ray fluxes using CCFs$^\dagger$.
 Column 8: Time bin size in s; 
 Column 9: represent fractional root mean square variability amplitude ($\rm F_{var}$) with errors and not defined when $S^2$ is lesser than $<\sigma^{2}_{err}>$ (for details see \S\ref{sec:lc_spec}).\\
$\dagger$: X-ray flux derived from spectral fitting are converted into luminosities using the distance to their corresponding clusters (see Table~\ref{tab:sample_info}). 
The spectral parameters are not derived for the  stars with poor count statistics and their unabsorbed X-ray fluxes have been estimated by their count rates in EPIC detector 
using CCFs (WebPIMMS), i.e.,  $\rm{Flux= CCF  \times Count rates}$.
The values of CCFs (in units of $\rm{erg~s^{-1}~cm^{2}}$) are derived for PN and MOS detectors.
For Intermediate mass stars :
 $\rm{3.926\times10^{-12}}$ and  $\rm{1.247\times10^{-11}}$  for NGC 869 at 2.07 keV ;   
 $\rm{3.528\times10^{-12}}$ and  $\rm{1.181\times10^{-11}}$  for NGC 884 at 1.30 keV ;  
$\rm {4.799\times10^{-12}}$ and  $\rm{1.433\times10^{-11}}$  for NGC 7380 at 2.38 keV ; 
$\rm {1.835\times10^{-11}}$ and  $\rm{6.740\times10^{-11}}$  for Hogg 15  at 0.29 keV; 
For Low mass stars : 
 $\rm{4.937\times10^{-12}}$ and  $\rm{1.847\times10^{-11}}$  for NGC 663 at 0.71 keV;   
 $\rm{3.926\times10^{-12}}$ and  $\rm{1.292\times10^{-11}}$  for NGC 869 at 2.05 keV;   
 $\rm{3.610\times10^{-12}}$ and  $\rm{1.206\times10^{-11}}$  for NGC 884 at 1.59 keV;  
$\rm {4.931\times10^{-12}}$ and  $\rm{1.460\times10^{-11}}$  for NGC 7380 at 2.80 keV ; 
$\rm {6.054\times10^{-12}}$ and  $\rm{1.749\times10^{-11}}$  for Berkeley 86 at 1.87 keV;     
$\rm {7.689\times10^{-11}}$ and  $\rm{2.290\times10^{-11}}$  for Hogg 15 at 0.97 keV ;   
}}
\end{list}
\end{deluxetable}

%% file: two_sample_test_new.tex
\begin{table*}
\caption{Results of two sample tests.}             
\label{tab:two_sample}      
\begin{tabular}{lll}       
\hline 
\multicolumn{3}{c}{Statistics of Objects in Two Groups}\\
\hline
Log($\rm{L_X}$) \egs                                  &  $>$ 30.8  \\
\hline
Number of stars (Low mass)                            &   100      \\
Number of stars (Intermediate mass)                   &   27      \\
\hline
\multicolumn{3}{c}{Probability of having a common parent $\rm{L_X}$ distribution}\\
\hline
Wilcoxon Rank Sum Test                   &    0.07 \\
Logrank Test                             &    0.02 \\
Peto and Peto Generalized Wilcoxon Test  &    0.08 \\
Kolmogorov-Smirnov Test                  &    0.23 \\
\hline
Log($\rm{L_X/L_{bol}}$)                  &       \\
\hline
\multicolumn{3}{c}{Probability of having a common parent $\rm{L_X/L_{bol}}$ distribution}\\
\hline
Wilcoxon Rank Sum Test                   &    8.1$\times10^{-13}$  \\
Logrank Test                             &    0.0            \\
Peto and Peto Generalized Wilcoxon Test  &    0.0           \\
Kolmogorov-Smirnov Test                  &    9.6$\times10^{-12}$  \\
\hline
\end{tabular}
\end{table*}

%% file: ms.bbl
\begin{thebibliography}{}

\bibitem{}Alexander F.,  Preibisch T., 2012, A\&A, 539, A64

\bibitem{}Baines D., Oudmaijer R. D., Porter J. M., Pozzo M.,2006,MNRAS,367,737  

\bibitem{}Balona  L. A., 2013, \mnras, 431,2240

\bibitem{}Baluci$\acute{n}$ska-Church M., McCammon D.,1992,\apj,400,699 




\bibitem{}Bhatt H., Pandey J. C., Kumar B., Sagar R., Singh K. P.,2010,New Astronomy,15,755
\bibitem{}Bhatt H., Pandey J. C., Singh K. P., Sagar Ram, B. Kumar, 2013, submitted to JAA 

\bibitem{}Bhavya B., Mathew B., Subramaniam A., 2007, BASI, 35, 383

\bibitem{}Brandt W. N., Hasinger G., 2005, ARA\&A, 43, 827

\bibitem{}Broos P. S., Feigelson E. D., Townsley L. K., Getman K. V., Wang J., Garmire G. P., Jiang Z., Tsuboi Y.,  2007, ApJS, 169, 353

\bibitem{}Bouvier J., Wichmann R., Grankin K., Allain S., Covino E., Fern$\acute{a}$ndez M., Martin E. L., Terranegra L., Catalano S., Marilli E., 1997, A\&A, 318,495 



\bibitem{}Caillault, J.-P.,  Helfand, D. J., 1985, ApJ, 289, 279
\bibitem{}Caramazza M., Micela G., Prisinzano L., Sciortino S., Damiani F., Favata F., Stauffer J. R., Vallenari A., Wolk S. J., 2012, \aa, 539, 74

\bibitem{}Chen W. P., Pandey A. K., Sharma Saurabh, Chen C. W., Chen Li, Sperauskas J., Ogura K., Chuang R. J., Boyle R. P., 2011, \aj, 142, 71

\bibitem{}Currie T., Evans N. R., Spitzbart B. D., Irwin J., Wolk S. J., Hernandez J., Kenyon S. J., 
      Pasachoff J. M.,2009,\aj,137,3210

\bibitem{}Currie T., Hernandez J., Irwin J., Kenyon S. J., Tokarz S., Balog Z., Bragg A., Berlind P., Calkins M.,  2010, ApJS, 186, 191

\bibitem{}Cutri R. M., Skrutskie M. F., van Dyk S., Beichman
 C. A., Carpenter J. M., Chester T., Cambresy L.,
 Evans T. et al., 2003, The IRSA 2MASS All-Sky Point
 Source Catalog, NASA/IPAC Infrared Science Archive
 http://irsa.ipac.caltech.edu/applications/Gator/

\bibitem{}Crowther P. A., 2007, A\&AR, 45, 177




\bibitem{}Daniel K. J., Linsky J. L., Gagn$\acute{e}$ M.,  2002, ApJ, 578, 486 

\bibitem{}Delgado A. J., Alfaro E. J., Yun J. L., 2007,\aa, 467,1397 

\bibitem{}Delgado A. J., Alfaro E. J., Yun J. L., 2011,\aa, 531,141  

\bibitem{}Dias W. S., Alessi B. S., Moitinho A., L$\acute{e}$pine J. R. D.,  2002, A\&A, 389, 871 

\bibitem{}Dobbie P. D., Lodieu N., Sharp R. G.,  2010, MNRAS, 409, 1002


\bibitem{}Donati J.-F., Semel M., Carter B. D., Rees D. E., Collier Cameron A.,1997,\mnras,291,658

\bibitem{}D'Orazi V., Randich S., 2009, A\&A, 501, 553 


\bibitem{}Edelson R., Krolik J., Pike G., 1990,\apj, 359, 86

\bibitem{}Edelson R., Turner T. J., Pounds K., Vaughan S., Markowitz A., Marshall H.,
   Dobbie P., Warwick R., 2002,\apj,568,610

	
\bibitem{}Fabricius C., Høg E., Makarov V. V., Mason B. D., Wycoff G. L., Urban S. E., 2002, A\&A, 384, 180



\bibitem{}Feigelson E. D., Broos P., Gaffney J. A. III, Garmire G., Hillenbrand L. A., Pravdo S. H., Townsley L., Tsuboi Y.,2002,\apj,574,258

\bibitem{}Feigelson E. D., Gaffney J. A. III, Garmire G., Hillenbrand L. A., Townsley L.,2003,\apj,584,911

\bibitem{}Feigelson, E. D., et al. 2004, ApJ, 611, 1107 
 


\bibitem{}Flaccomio E., Micela G., Sciortino S.,2003,\aa,402,277 

\bibitem{}Flaccomio E., Damiani F., Micela G., Sciortino S.,
      Harnden F. R. Jr., Murray S. S., Wolk S. J.,2003,\apj,582,398

\bibitem{}Flesch E., 2010, PASA, 27, 283	
	 









\bibitem{}Getman K. V., Feigelson E. D., Townsley L., Bally J., Lada C. J., Reipurth B., 2002, \apj, 575, 354



\bibitem{}Glebocki R., Gnacinski P., 2005, Catalog of Stellar Rotational Velocities, 3244	

\bibitem{}Girardi L., Bertelli G., Bressan A., Chiosi C., Groenewegen M. A. T., Marigo P., Salasnich B., Weiss A., 2002, A\&A, 391, 195



\bibitem{}Guarcello M. G., Caramazza M., Micela G., Sciortino S., Drake J. J., Prisinzano L., 2012, \apj, 753, 117

\bibitem{}G$\rm{\ddot{u}}$del M., 2004,  A\&A Rev., 12, 71




\bibitem{}Hubrig S., Sch$\ddot{o}$ller M., Yudin R.V., 2004, \aa, 428, 1 

\bibitem{} Ikeda H., et al. 2008, AJ, 135, 2323


\bibitem{}Jeffries R. D., Evans P. A., Pye J. P., Briggs K. R.,2006,\mnras,367,781

\bibitem{}Joshi H., Kumar B., Singh K. P., Sagar R., Sharma S., Pandey, J. C., 2008, \mnras, 391, 1279 

\bibitem{}Kazarovets E. V., Samus N. N., Durlevich O. V., 2001, Information Bulletin on Variable Stars, 5135, 1

\bibitem{}Kudritzki R. P., Puls J., 2000, ARA\&A, 38, 613




\bibitem{}Kharchenko N. V., Roeser S., 2009,  1280 , Available at VizieR On-line Data Catalog http://webviz.u-strasbg.fr/viz-bin/VizieR






\bibitem{}La Palombara N., Mereghetti S., 2006, A\&A, 455, 283

\bibitem{}Lopes de Oliveira R., Motch C., Smith M. A., Negueruela I., Torrej$\acute{o}$n J. M.,  2007, A\&A, 474, 983 

\bibitem{}Lucy L. B., White R. L., 1980, ApJ, 241, 300
 


 
    

\bibitem{}Micela G., Sciortino S., Favata F., Pallavicini R., Pye J., 1999, A\&A, 344, 83

\bibitem{}Micela, G., Sciortino, S., Serio, S., et al. 1985, ApJ, 292, 172 
\bibitem{}Micela, G., Sciortino, S., Vaiana, G. S., et al. 1988, ApJ, 325, 798 
\bibitem{}Micela, G., Sciortino, S., Vaiana, G. S., et al. 1990, ApJ, 348, 557 
 Solar flares: A monograph from Skylab Solar Workshop II  (A80-37026 15-92), 341-409



\bibitem{}Naz$\acute{e}$ Y., 2009, A\&A, 506, 1055


\bibitem{}Ogura K., Sugitani K., Pickles A., 2002, AJ, 123, 2597

\bibitem{}Osten R. A., et al., 2010, ApJ, 721, 785

\bibitem{}Owocki S. P., Cohen D. H., 1999, ApJ, 520, 833 

\bibitem{}Pandey A. K., Upadhyay K., Ogura K., Sagar R., Mohan V., Mito H., Bhatt H. C., Bhatt B. C., 2005, \mnras, 358, 1290




\bibitem{}Patel M. K., Pandey J. C., Savanov I. S., Prasad V., Srivastava D. C., 2013, \mnras, 430, 2154 	

\bibitem{}Patten B. M., Simon T., 1996, ApJS, 106, 489





\bibitem{}Pickles A., Depagne E., 2010, PASP, 122,1437
Feigelson, E. D., et al. 2004, ApJ, 611, 1107
\bibitem{}Pizzolato N., Maggio A., Micela G., Sciortino S., Ventura P., D'Antona F. , 2000, Proceedings from ASP Conference, Vol. 198., 
ed. by R. Pallavicini, G. Micela, and S. Sciortino ISBN: 1-58381-025-0, 71


\bibitem{}Press, W. H., Teukolsky, S. A., Vetterling, W. T., et al. 1992, Numerical recipes in FORTRAN. The art of scientific computing.
 

\bibitem{}Preibisch T., 1997, A\&A, 324, 690 

\bibitem{}Preibisch T., Kim Y.-C., Favata F., Feigelson E. D., Flaccomio E., Getman K., Micela G., Sciortino S. et al.,2005,ApJS,160,401

\bibitem{}Preibisch T., Feigelson E. D., 2005, ApJS, 160, 390
 






\bibitem{}Sagar R., Munari U., de Boer K. S., 2001,\mnras,327,23 




\bibitem{}Scholz A., Coffey J., Brandeker A., Jayawardhana R., 2007, ApJ, 662, 1254


\bibitem{}Siess L., Dufour E., Forestini M., 2000, \aa, 358, 593

\bibitem{}Slesnick C. L., Hillenbrand L. A., Massey P.,2002,\apj,576,880




\bibitem{}Skiff B. A., 2007, yCat, 102023S,  Available at VizieR On-line Data Catalog http://webviz.u-strasbg.fr/viz-bin/VizieR

\bibitem{}Smith R. K., Brickhouse N. S., Liedahl D. A., Raymond J. C., 2001, \apj, 595, 365

\bibitem{}Stassun K. G., Ardila D. R., Barsony M., Basri G., Mathieu R. D.,2004,\aj,127,3537 

 


 
\bibitem{}Stelzer B., Hu$\rm{\acute{e}}$lamo N., Micela G., Hubrig S., 2006,\aa,452,1001  
\bibitem{}Stern, R. A., Zolcinski, M. C., Antiochos, S. K., Underwood, J. H., 1981, ApJ, 249, 647
\bibitem{}Strom S.E., Wolff S.C., Dror D.H.A., 2005, AJ, 129, 809 

\bibitem{}Str$\ddot{u}$der L. et al. 2001, A\&A, 365, 18



\bibitem{}Turner M. J. L. et al. 2001, A\&A, 365, 27



\bibitem{}Vaiana, G. S., et al. 1981, ApJ, 245, 163

\bibitem{}Vuong M. H., Montmerle T., Grosso N., Feigelson E. D., Verstraete L., Ozawa H., 2003, \aa, 408, 581

\bibitem{}Wade G. A., Drouin D., Bagnulo S., Landstreet J. D.,
          Mason E., Silvester J., Alecian E.,
          B$\rm{\ddot{o}}$hm T. et al.,2005,\aa,442,31

\bibitem{}Wright N. J., Drake J. J., Civano F., 2010, \apj, 725, 480



\bibitem{}Zhekov S. A., Palla Francesco, 2007, \mnras, 382, 1124 



\end{thebibliography}
